\documentstyle[aps,epsfig]{revtex}

\newcommand{\beq}{\begin{equation}}
\newcommand{\eeq}{\end{equation}}
\newcommand{\beqa}{\begin{eqnarray}}
\newcommand{\eeqa}{\end{eqnarray}}

\def\Tsp{T_{\mathrm \scriptsize sp}}   
\def\tsp{t_{\mathrm\scriptsize sp}}
\def\Tm{T_m}

\begin{document}

\title{Glass and polycrystal states in a lattice spin model}

\author{Andrea Cavagna, Irene Giardina and Tom\'as S.~Grigera}

\address{ Center for Statistical Mechanics and Complexity, INFM Roma
``La Sapienza'' and} 
\address{Dipartimento di Fisica, Universit\`a di Roma ``La Sapienza'',
00185 Roma, Italy}
             
\date{October 17, 2002}

\maketitle

\begin{abstract}
We numerically study a nondisordered lattice spin system with a 
first order liquid-crystal transition, as a model for supercooled 
liquids and glasses. Below the melting temperature the system
can be kept in the metastable liquid phase, and it displays a 
dynamic phenomenology analogous to fragile supercooled liquids, 
with stretched exponential relaxation, power law increase of 
the relaxation time and high fragility index. 
At an effective spinodal temperature $\Tsp$ the relaxation time 
exceeds the crystal nucleation time, and the supercooled liquid 
loses stability. Below $\Tsp$ liquid properties cannot be extrapolated, 
in line with Kauzmann's scenario of a `lower metastability limit' of 
supercooled liquids as a solution of Kauzmann's paradox. 
The off-equilibrium dynamics below $\Tsp$ corresponds to fast 
nucleation of small, but stable, crystal droplets, followed by 
extremely slow growth, due to the presence of pinning energy barriers.
In the early time region, which is longer the lower the temperature, this 
crystal-growth phase is indistinguishable from an off-equilibrium glass, 
both from a structural and a dynamical point of view: crystal growth has 
not advanced enough to be structurally detectable, and a violation of 
the fluctuation-dissipation theorem (FDT) typical of structural glasses 
is observed. On the other hand, for longer times crystallization reaches 
a threshold beyond which crystal domains are easily identified, and FDT 
violation becomes compatible with ordinary domain growth.
\end{abstract}

\section{Introduction}

When a liquid is cooled fast enough below its melting point,
crystallization is avoided and the system enters the supercooled
phase.  Relaxation time increases rapidly in this temperature regime,
and when it becomes comparable with the largest experimentally
accessible time the system falls out of equilibrium, remaining stuck
in a disordered phase called {\em structural glass} \cite{vetri}.

The configurational disorder of a structural glass is not caused by
the presence of intrinsic disorder in the Hamiltonian, since this is
just the sum of deterministic interaction elements.  This
notwithstanding, a lot of attention has been devoted to the
phenomenological analogies between structural glasses and mean-field
spin-glasses, which are systems directly containing quenched disorder
in the Hamiltonian in the form of random couplings among the spins
\cite{spin-glass}.  In particular, models of spin-glasses with
$p$-body interactions and $p\geq 3$, seem indeed to have many features
in common with structural glasses \cite{thirumalai}, with the obvious
(but crucial) difference that in mean-field systems there is a purely
dynamical transition which is absent in real finite-dimensional
glasses.  Despite the close phenomenology, however, it is natural to
ask why we use models with quenched disorder to understand the
behaviour of physical systems with deterministic Hamiltonians.

Disorder, one of the characteristics of the glassy state, cannot be
expected to arise spontaneously from a deterministic mean-field
Hamiltonian, because in this case the mechanically stable
configurations are ordered: the local field is the same for all spins,
and thus all tend to point in the same direction.  Thus if we wish to
stick to mean-field and the associated analytical advantages, we must
resort to quenched disorder to produce glassy behaviour.  The argument
does not apply to finite range systems, however, and we can ask
whether finite dimensional spin systems with no quenched disorder and
$p\geq 3$ display a glassy phenomenology similar to structural glasses
(we of course know that $p=2$, i.e.\ the Ising model, is not
glassy under many respects). This is a relevant question, first because 
it is important to understand what the key ingredients responsible for 
glassy dynamics are, and second because lattice spin systems are normally 
easier to study than off-lattice liquid models.

The answer to this question seems to be affirmative. A number of spin models without
quenched disorder now exist that reproduce a phenomenology similar to
supercooled liquids and glasses. Generally, one can say that in all
these systems the non-trivial dynamic behaviour is related to the
presence of frustration, due to complex even if non-disordered
interaction terms in the Hamiltonian.  In the SS model of
ref.~\cite{sethna}, this is achieved by mixing nearest- and
next-nearest-neighbor interactions, which gives an interesting
phenomenology, although not precisely glassy.  In the plaquette (PQ)
model of \cite{lipo,lipo-bray,juanpe-plaq}, on the other hand, glassiness is
achieved by a $p$-body interaction term with $p=4$, in close analogy
with $p$-spin models of mean-field spin-glasses. In the lattice glass
model of \cite{meza} frustration arises from a constraint on the 
numbers of neighboring particles on the lattice. The PQ model has been
heavily studied in recent years, and the results seem to suggest that
its phenomenology has at least some points in common with structural
glasses. In particular, a sufficiently rapid cooling of the PQ model
forms an off-equilibrium disordered phase which closely resembles real
glasses. Therefore, it would seem that $p$-body interactions in finite
dimensional spin models without quenched disorder are sufficient to
reproduce a phenomenology similar to supercooled liquids and glasses.

The aim of this paper is to study a model in the same class as those 
just described, namely
the deterministic version of the CTLS model introduced in
\cite{tomas-ctls} (a concise description of some of our results has recently
appeared in \cite{minimanga}). As discussed below, this model has various
advantages with respect to its predecessors
and will enable us to check more deeply to what extent lattice spin
systems can mimic structural glasses. Moreover, important questions 
on the low temperature behaviour of glasses, such as the stability of 
the supercooled phase and the role of the crystal, can be addressed, 
whereas in previous models such an analysis was very limited. Let us 
now explain why.

Liquids in nature have a crystalline ground state, usually with low
degeneracy, and display a first order thermodynamic transition at the
melting temperature. In the PQ model in three dimensions a first order
transition takes place, but the `crystalline' ground state has an
unusual property for liquids, in that it is strongly degenerate: for
linear size $L$, the PQ model has $2^{3L}$ different ground states
with the same energy. The problem is not the degeneracy in itself
(note that the ground state entropy density is zero), but rather the
consequences that this degeneracy has on the possibility to measure
the degree of order in the system. More precisely, in the PQ model,
given a ground state configuration (for example the ferromagnetic
one), we can obtain another ground state by flipping any plane of
spins. Clearly, by iterating this procedure we can produce ground
state configurations which look highly disordered, despite all having
the ground state energy. The problem then is how to distinguish a
truly disordered, or glassy, configuration, from a configuration made
up of many droplets of different ground states.  The energy of both
configurations will be high, in the first case because it is
essentially a liquid, in the second case because of the interfaces
among different droplets. In other words, energy is not a sufficiently
precise measure of order in the PQ model, and due to the strong ground
state degeneracy no configurational means can be devised to quantify
order and domain growth.

The system we introduce here does not suffer from this problem: it has
a doubly degenerate ground state, such that it is possible to measure
the amount of crystalline order and to monitor domain growth.  A
likely objection is that glassy physics does not have anything to do
with the crystal, and that the possibility to measure the formation
and growth of crystal domains is far from essential to study the
physics of glasses. We reply to this objection with two observations.
First, that simple glass formers usually {\em do have} a crystalline
phase, whose possible connection with the properties of the glass
cannot be completely ruled out in principle. In fact, our model explicitly
shows such a connection. Secondly, if on the other hand spin models with
a crystal phase fail somehow to reproduce some basic property of
glasses, then we should understand why, and make sure that all spin
models are free from this problem. Indeed, the precise nature of what
is normally called the 'glassy' phase in systems like PQ \cite{lipo}
and in lattice glass models \cite{meza} is the main focus of our work.

Our model shows, in this respect, an interesting phenomenology.  As we
will show, it has a first order liquid-solid transition at a melting
temperature $T_m$. Crystallization can be avoided by fast cooling, and
a supercooled liquid phase, metastable with respect to the crystal, is
found.  ``Equilibrium'' measurements can be performed as long as the
equilibration time of the supercooled liquid is much shorter than the
nucleation time of the crystal.  In this phase the behaviour of our
system is very similar to real supercooled liquids, and in particular
to fragile systems.  However, a temperature $\Tsp$ exists, where the crystal
nucleation time becomes of the same order as the liquid equilibration
time: at this point the supercooled liquid looses stability and it no
longer exists below $\Tsp$. We call $\Tsp$ the {\em effective
spinodal} temperature. 
It is important to stress the following fact: we
are able to detect the loss of stability of the liquid only because the
equilibration time at $\Tsp$, $\tsp$, is shorter than our longest
experimental time. We can state this point in a different way. A
system falls out of equilibrium when the relaxation time becomes too
large for the (real/numerical) experiments, and this is the operational
definition of the glass transition temperature $T_g$.  Therefore, the
reason why we can observe the loss of stability of the supercooled
liquid in our model is that $\Tsp > T_g$.  The Kauzmann paradox
(i.e.\ the fact that the extrapolated liquid entropy becomes smaller
than the crystal entropy at a finite temperature) is avoided by this
loss of stability: no equilibrium supercooled liquid exists below
$\Tsp$ and the entropy extrapolation is meaningless. In fact, this is
the scenario proposed by Kauzmann himself \cite{kauzmann} as solution
of the paradox, which is thus exactly reproduced in this system.
Below $\Tsp$ the system enters a completely different phase, where
crystal nucleation and crystalline domain growth are the main dynamic
processes. However, by quenching to lower temperatures the growth of
crystalline order becomes so slow that it is impossible to distinguish
this phase from a typical off-equilibrium realistic glass. Moreover,
we shall show that dynamical measurements of the
fluctuation-dissipation (FD) ratio are compatible with the results
obtained in ordinary structural glasses below $T_g$.

It is our impression that the scenario described above is also shared
by the PQ, and possibly other spin models. In particular, that what is
normally called the glassy phase for the PQ model is an
off-equilibrium regime of very slow crystal growth. However, as
already mentioned, without a direct measure of crystalline order it is
hard to assess the amount of order in the PQ model, so the question
whether there is in fact a glassy phase qualitatively different from
the crystal growth phase below $\Tsp$ is open. This is certainly not
the case in the present model, though.

Throughout the paper we will use a notation common to supercooled
liquids and glasses. We call $T_m$ the melting temperature.  In
analogy with mode coupling theory \cite{mct}, $T_c$ will be the
temperature where a power law fit locates a divergence of the
relaxation time, while $T_0$ will be the temperature where a
Vogel-Fulchner-Tamman fit puts the same divergence.  The so-called
Kauzmann temperature, i.e. the temperature where the extrapolated
liquid entropy becomes equal to the crystal entropy, is called $T_s$.
The effective spinodal temperature, i.e.\ the temperature below which
it becomes impossible to observe the metastable liquid is called
$\Tsp$.  This temperature is called $T_g$ in \cite{lipo}, but we
reject this choice because, as already mentioned, in the liquid
literature the glass transition is defined rather as the point where
the equilibration time is longer than the experimental time.

The paper is structured as follows.  In Section II we introduce our
model, and describe the properties of the crystalline ground state. In
Section III we show, by comparing the liquid and crystal free
energies, that the model has a first order phase transition at a
melting temperature $T_m$.  The dynamics of the supercooled liquid
below $T_m$ and the metastability limit $\Tsp$ are considered in
Section IV, while aging dynamics and its relation to domain growth are
studied in Section V. Finally, we draw our conclusions in Section VI.

\section{The Model}

The model is described by the Hamiltonian
\begin{equation}
H=\sum_{i=1}^N f_i (1+s_i)  ,  \label{hamiltonian}
\end{equation}
where the variables $s_i=\pm 1$ are Ising spins belonging to a
two-dimensional square lattice of side $L$, and the plaquette variable
$f_i$ is the product of the four nearest neighbors of spin $i$:
\begin{equation}
f_i=s_i^{W}s_i^{S}s_i^{E}s_i^{N} \ ,
\end{equation}
($W$ is for west, $S$ for south and so on). The model is studied by
Metropolis Monte Carlo simulations with single-spin-flip dynamics for
lattices of size $L=100$ and $L=500$.

A disordered version of this model was first introduced in ref.\
\cite{tomas-ctls}, as a way to model two-level systems in a lattice,
interacting with their nearest neighbors, whence its name CTLS
(Coupled Two Level Systems). In fact, one can take the spin $s_i$ to
be the variable describing the state of a two-level system sitting at
site $i$. Then state $s_i=-1$ has always zero energy, while the other
state can be at 1 or -1 depending on the factor $f_i$. Thus when one
of the neighbors switches, the relative height of the two states at
$i$ is reversed.  In \cite{tomas-ctls}, however, quenched random
coupling constants where considered. Here we shall see that the
multi-body nature of the interaction is sufficient to give a
non-trivial dynamic and static behaviour.

\begin{figure}
\begin{center}
\leavevmode
\epsfxsize=4in,
\epsffile{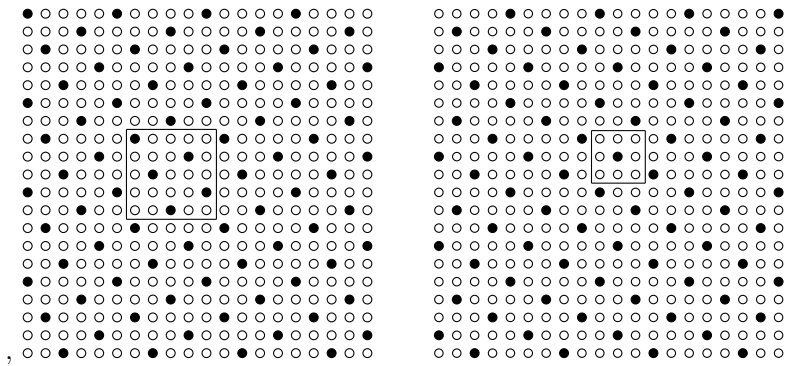}
\vglue 0.2 truecm
\caption{Dextrocrystal (left) and levocrystal (right): empty circles
correspond to positive spins, full circles to negative spins.  In the
left figure the square marks the five-element unit cell. In the
right figure the small square shows the elementary unit used in the
definition of the crystal mass (see sect.~\ref{sec-thermo}).}
\label{ground-state}
\end{center}
\end{figure}

Frustration in the CTLS comes from the fact that it is not possible to
satisfy simultaneously the 4-spin and the 5-spin interaction. After
careful inspection of the Hamiltonian it is found that the ground
state of the CTLS is obtained by covering the lattice with the
following non-overlapping elements:
\begin{equation}
s_i=-1, \qquad s_i^{W}=s_i^{S}=s_i^{E}=s_i^{N}=+1 \  . \label{element}
\end{equation}
As Fig.~\ref{ground-state} shows, the covering can be done in two
ways, giving rise to two different ground states which we call
dextrocrystal and levocrystal. These two crystals are connected by the
symmetry $x\to -x, y\to y$, or $x\to x, y\to -y$, while they are both
invariant under $x\to -x, y\to -y$. Therefore, the CTLS has a
crystalline ground state, with energy density $E_{GS}=-1.6$.  In
Fig.~\ref{ground-state} it can be seen that the crystal can be
obtained by periodic repetition of a $5\times 5$ unit cell.  Apart
from the two distinct dextro- and levo- forms, additional crystals can
be formed by translating the unit cell up to $5$ lattice spacings in each
direction.  There
are then different ways to cover the lattice, which are not locally
compatible with each other. Upon a quench, the system will greedily
optimize the energy locally in an uncoordinated way, creating many
different competing domains with boundaries that turn out to be
pinned, thus giving rise to slow dynamics.

As already anticipated, the low degeneracy of the
crystalline ground state is one of the main advantages of the CTLS
compared to other plaquette models \cite{lipo}.  Indeed, as we shall
see, in the CTLS it is possible to quantify the amount of crystalline
order in the system by purely configurational means, irrespective of
the value of the energy. This is particularly useful when we are
studying the growth of crystalline domains, because in this case the
energy may be very large, due to interfaces, even though there is a
substantial amount of crystalline order. Moreover, the low degeneracy
of the crystal makes this model more similar to real liquids.

\section{Thermodynamics}

\label{sec-thermo}

\begin{figure}
\begin{center}
\leavevmode
\epsfxsize=3.5in
\epsffile{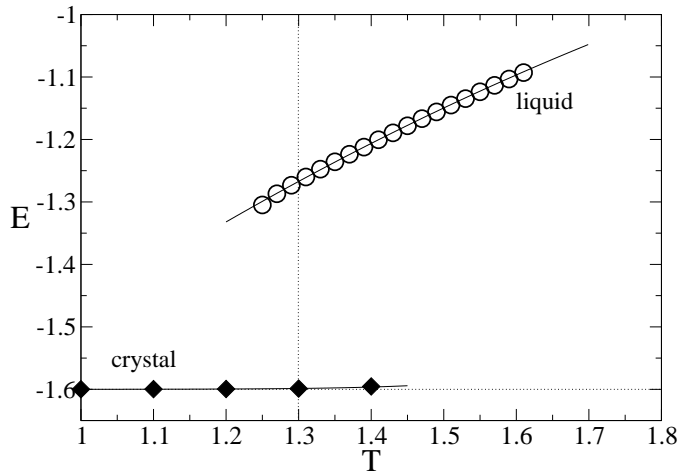}
\vglue 0.2 truecm \caption{Liquid and crystal energy per spin as a function of
temperature. The continuous lines correspond to the two fits reported
in the text. The horizontal dotted line marks the energy of the crystal at zero
temperature $E_{GS}=-1.6$. The vertical dotted lines marks $T_m=1.30$. $L=100$.}
\label{metastability}
\end{center}
\end{figure}

In the present section we will show that the CTLS has a first order
phase transition at a melting temperature $T_m$.  To prove this fact
we start from the numerical observation that in a certain interval of
temperature both the crystalline phase obtained by heating the ground
state and the liquid phase obtained by cooling a high temperature
configuration are stable up to times of order $10^6$ Montecarlo steps
(MCS), for $L=100$. In Fig.~\ref{metastability} we plot the
equilibrium internal energy as a function of the temperature for the
crystal and the liquid phases.

In order to find $T_m$ we need to compute the free energy in the two
phases by integration of the internal energy.  To this aim, we fit the
liquid energy (LQ) as
\begin{equation}
E_{LQ}(T)= -a\tanh (b/T) , \qquad a=1.84, \quad b=1.10,
\end{equation}
while for the crystal (CR) we find
\begin{equation}
E_{CR}(T)= E_{GS} + c T^n , \qquad c=5.5\times10^{-5}, \quad n=13 .
\end{equation}
By using the thermodynamic formula,
\begin{equation}
\beta F(\beta)=\beta_0 F(\beta_0) + \int_{\beta_0}^\beta d\beta'
E(\beta') , \qquad \beta=1/T,
\end{equation}
we obtain the free energy of the liquid and crystal phases:
\begin{eqnarray}
F_{LQ}(T)&=& -T \log 2- \frac{a}{b}\,T\log\cosh(b/T) , \\
F_{CR}(T)&=& E_{GS} - \frac{c}{n-1} T^n  , 
\end{eqnarray}
where we have taken $\beta_0=0$ for the liquid, and $\beta_0=\infty$
for the crystal.  The melting temperature is fixed by requiring
$F_{LQ}(T_m)=F_{CR}(T_m)$, which gives
\begin{equation}
 T_m=1.30 .
\end{equation} 
We have checked that our energy data are independent of system size
for $100 \leq L \leq 500$.

\begin{figure}
\begin{center}
\leavevmode
\epsfxsize=3in
\epsffile{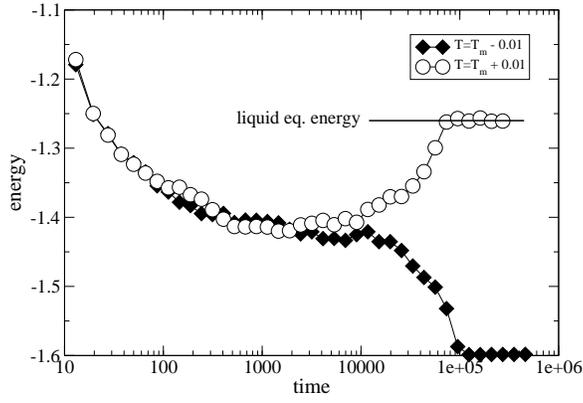}
\vglue 0.2 truecm \caption{Energy per spin as a function of time below and above $T_m$. $L=100$.}
\label{meta-ene}
\end{center}
\end{figure}

In order to check that a first order liquid-crystal transition
actually takes place at $T_m$ we proceed as in \cite{lipo}: we
consider an initial condition where half of the system is in the
crystal phase and half in a random phase, and we run this
configuration at $T_m+0.01$ and at $T_m-0.01$. If our determination of
$T_m$ is correct, after a transient time the system should relax
either to the liquid or crystalline phase, since the stable phase will
expand and take over the whole system after a sufficiently long
time. The time evolution of the energy is shown in
Fig.~\ref{meta-ene}, confirming our previous determination of
$T_m$. We therefore conclude that the CTLS has a first order
thermodynamic phase transition at $T_m=1.30$.

\begin{figure}
\begin{center}
\leavevmode
\epsfxsize=3in
\epsffile{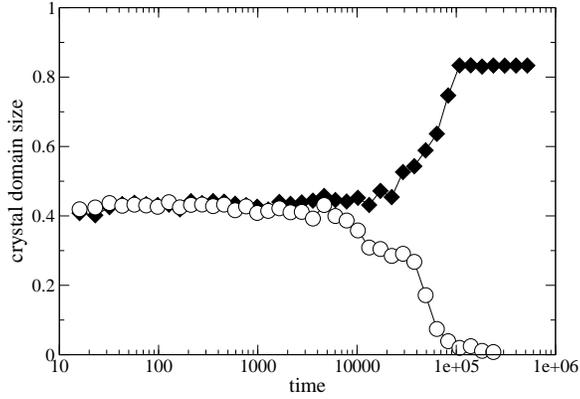}
\vglue 0.2 truecm \caption{Domain size $\xi$ as a function of time below and above $T_m$. Same
temperatures and symbols as in Fig.~\ref{meta-ene}.}
\label{meta-xi}
\end{center}
\end{figure}

\begin{figure}
\begin{center}
\leavevmode
\epsfxsize=3in
\epsffile{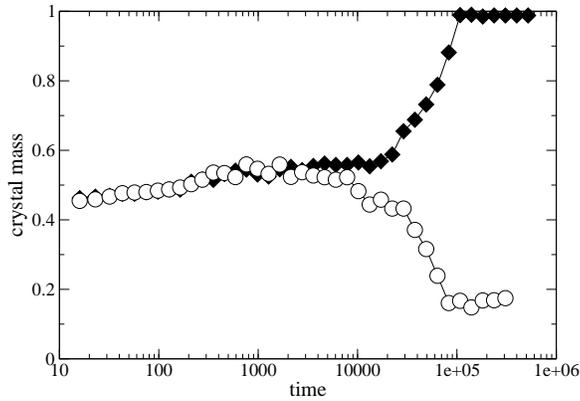}
\vglue 0.2 truecm \caption{Crystal mass $m$  as a function of time below and above $T_m$. Same
temperatures and symbols as in Fig.~\ref{meta-ene}.}
\label{meta-mass}
\end{center}
\end{figure}

Even though the energy of a configuration may be considered as an
index of the degree of crystallization of the system, it is certainly
not a very sensitive one. Given that we know the crystal in this
system, and thanks to its low degeneracy, we introduce some methods to
directly quantify the amount of crystalline order in our samples.
This can be done either by measuring the typical crystal droplet size
$\xi$, or by measuring the total crystal mass $m$.  To measure $\xi$
we observe that both crystals have a periodic structure with a unit
cell of size $5$.  Thus, the Fourier transform of the spatial
correlation function,
\begin{equation}
G(r)=\frac{ \langle s(0)\, s(r) \rangle - \langle s\rangle ^2}
{1-\langle s\rangle ^2} \ ,
\end{equation}
has a peak at $k_0=2\pi/5$. The width $\lambda$ of this peak is
proportional to the inverse of the typical crystalline domain size.
We want to extract from $\lambda$ a measure of how large is the typical 
crystal domain, a question which is better answered by comparing it to the
system size. So we actually define $\xi$ as a density,
i.e. $\xi=1/(\lambda L)\in [0:1]$.  It is important to note that this
definition of $\xi$ has a limited power of resolution: in a random
configuration we find on average that $\xi_0=0.05$, which can
therefore be considered as the effective zero for $\xi$.  When we
think of a configuration entirely made up of tiny mismatched
crystallites we realize that $\xi$ is not enough as a measure of
crystal order: in this case $\xi$ would be very small, even though the
total amount of crystal would be large. The crystalline mass $m$ can
be measured by counting the fraction of crystallized spins. To this
aim, we have to define an elementary unit, which is large enough to
have a small probability to be formed randomly, but small enough to be
sensitive to small amounts of crystal. We choose the nine-spin unit
shown in Fig.~\ref{ground-state}(right): in the pure crystal each minus spin
is surrounded by eight pluses, and in the unit cell we have 5 of these
elements (which clearly overlap). To count all the spins in the cell
we must multiply the number of these units by 5. Thus, we define the
crystal mass density $m\in [0:1]$ as the number of these elementary units times
$5$ normalized by $L^2$. As for $\xi$, we have a random average value
of the mass, $m_0=0.01$, which is the effective zero of $m$.  Had we
chosen the $5\times 5$ unit cell, the accuracy would be higher, 
but the sensitivity would decrease as well.  In order
to test the definition of $\xi$ and $m$, we measure their time
evolution in the experiment described above for the determination of
$T_m$. This results are shown in Fig.~\ref{meta-xi} and
Fig.~\ref{meta-mass}.  Note that in the liquid phase at $T=T_m+0.01$
we have $m\sim 0.15 \gg m_0$ (whereas $\xi$ is compatible with zero), meaning
that also in the liquid above $T_m$ there is a certain amount of short
range order. From this experiment we see that both these quantities
are excellent configurational indicators of the amount of crystalline
order in the system.

We finish this section by noting that if we extrapolate $E_{LQ}(T)$
and $F_{LQ}(T)$ to low temperatures, we can find the temperature $T_s$
where the entropy of the supercooled liquid becomes equal to the
entropy of the crystal (which, in this temperature range, is to a very
good degree of approximation equal to zero). We find
\begin{equation}
T_s=0.91 .
\label{Ts}
\end{equation}
Therefore, $T_s$ is the temperature where the Kauzmann paradox occurs
\cite{kauzmann}: below $T_s$ the supercooled liquid entropy would
become smaller than that of the crystal, that is, quite unacceptably,
smaller than zero.  This temperature is usually called Kauzmann
temperature in the literature, a denomination we find somewhat
unsatisfactory given that Kauzmann himself did not believe the
extrapolation was really meaningful. Indeed, as we shall see, the
Kauzmann paradox is avoided in the CTLS.

\section{Dynamics}

The aim of this section is to study the dynamical behaviour of the
CTLS in the supercooled liquid phase (i.e.\ for $T<T_m$) and in
particular to analyze how the relaxation time increases when
decreasing the temperature.  As already anticipated in the
Introduction, there is a marked difference in the behaviour of the
system if our observational time window is longer or shorter than a
given time $t_{\mathrm \scriptsize sp}\sim 10^4$ MCS. For
$t<t_{\mathrm \scriptsize sp}$ the supercooled liquid is stable at all
temperatures where equilibration may be achieved: in this time regime
we can either study the equilibrium supercooled liquid or the
off-equilibrium glass, formed by a sufficiently fast cooling.  On the
other hand, if we are able to observe the dynamics on time scales
$t>t_{\mathrm \scriptsize sp}$ we discover that the supercooled liquid
loses stability at low enough temperatures, and the system enters a
completely different regime. Therefore, it is useful to divide our
dynamical investigation into two parts: first we shall assume that our
maximum experimental time $t_{\mathrm exp}$ is shorter than $t_{\mathrm
\scriptsize sp}$ and we will measure the equilibrium properties of the
supercooled liquid.  Second, we will consider times larger than
$t_{\mathrm \scriptsize sp}$, to analyze the loss of stability of the
supercooled liquid.  The scenario that will emerge from this analysis
is the following. Our system, if analyzed on short time scales
exhibits all the features of typical fragile structural glasses,
posing the usual questions about the nature of the glassy state at low
temperature. Only when looking at longer time-scales it is possible to
unveil the behaviour of the system and give explanations which would
otherwise be out of reach.

\subsection{Moderate times dynamics: $t<t_{\mathrm \scriptsize sp}$}

In this sub-section we will assume that the longest experimental time
available to us is $t_{\mathrm exp}\leq t_{\mathrm \scriptsize
sp}\sim10^4$ MCS.  To probe the dynamics of the CTLS we measure the
equilibrium relaxation time $\tau$ as a function of the temperature
$T$.  This can be obtained from the decay rate of the (normalized)
equilibrium spin-spin correlation function, defined as,
\begin{equation}
C(t,t_w)=\frac{ \langle s(t_w)\, s(t_w+t) \rangle  - \langle s\rangle^2}
{1-\langle s\rangle^2} \ ,
\end{equation}
where $\langle \cdots \rangle$ indicate averages over spins and
initial configurations.  We quench the system from infinite
temperature down to the target temperature $T$ and wait for it to
equilibrate. At equilibrium the dependence on $t_w$ disappears and
$C=C(t)$. This is our first equilibration test.  In order to extract a
relaxation time from the correlation function we fit it with a
stretched exponential, which is the expected equilibrium behaviour for
$C(t)$ in super-cooled liquids
\begin{equation}
C(t)=\exp\left[-(t/\tau)^\beta\right]  .
\label{stretched}
\end{equation} 

\begin{figure}
\begin{center}
\leavevmode
\epsfxsize=3.5 in
\epsffile{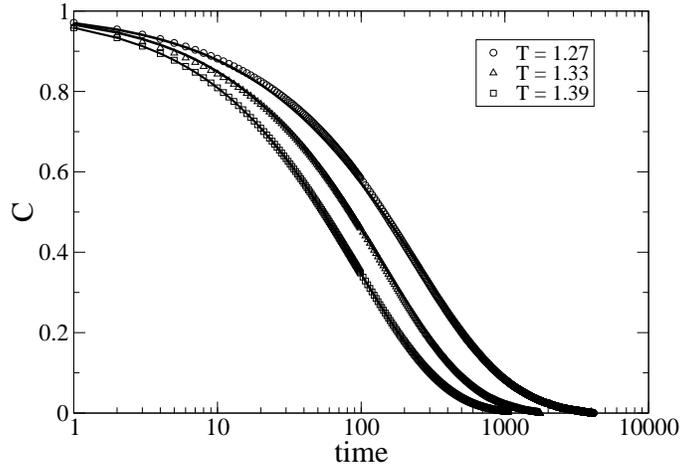}
\vglue 0.2 truecm \caption{Correlation function vs time for three different temperatures
(symbols) and stretched exponential fits (continuous lines), in the 
supercooled phase $T<T_m$. $L=500$.}
\label{corre}
\end{center}
\end{figure}

In Fig.~\ref{corre} we plot $C(t)$ for three different temperatures,
together with their stretched exponential fit.  To be sure that the
system has equilibrated we check that at each temperature the
equilibration time that we have waited is much larger than the
relaxation time $\tau$, as obtained from the fits (\ref{stretched}).
More precisely, since $C(t)$ decays to zero in approximately $20\tau$
we require $t_{eq}\geq 20 \tau$.

\begin{figure}
\begin{center}
\leavevmode
\epsfxsize=3.5 in
\epsffile{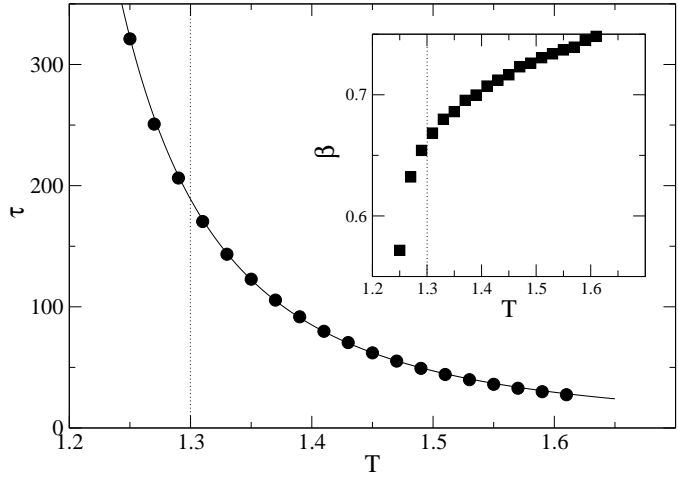}
\vglue 0.2 truecm \caption{Relaxation time as a function of temperature. The full line 
is the power law fit eq.(\ref{power}). Inset:
Kohlrausch $\beta$ exponent of the stretched exponential fit vs.\
temperature. $L=500$.}
\label{relax}
\end{center}
\end{figure}

The relaxation time as a function of the temperature is plotted in
Fig.~\ref{relax}. We note that there is no dynamic signature of the
melting transition $T_m$. Below $T_m$ the supercooled liquid is
metastable with respect to the crystal, and there is a sharp increase
of the relaxation time. In the inset of Fig.~\ref{relax} we can see
that the Kohlrausch $\beta$ exponent of the stretched exponential fit
decreases with the temperature, as it happens for realistic models of
liquids (see, for example, the data of \cite{sastry}).

In the temperature regime of Fig.~\ref{relax} 
the relaxation time can be fitted very well by a power law,
\begin{equation}
 \tau=\frac{A}{(T-T_c)^\gamma} , \qquad \mbox{with} \quad
 T_c=1.06, \quad \gamma=2.29 .
\label{power}
\end{equation}
This fact suggests that the CTLS is a fragile system \cite{angell}.
To support this conclusion, we compare the CTLS with an
Arrhenius/strong system. In \cite{juanpe-plaq} it has been noted that
the PQ model in $d=2$ is such a system, and we therefore compare the
two models by plotting the logarithm of $\tau$ as a function of
$T^*/T$, where $T^*$ is an normalization temperature, arbitrarily 
defined as the point where $\tau(T^*)=1000$.
\begin{figure}
\begin{center}
\leavevmode
\epsfxsize=3.5 in
\epsffile{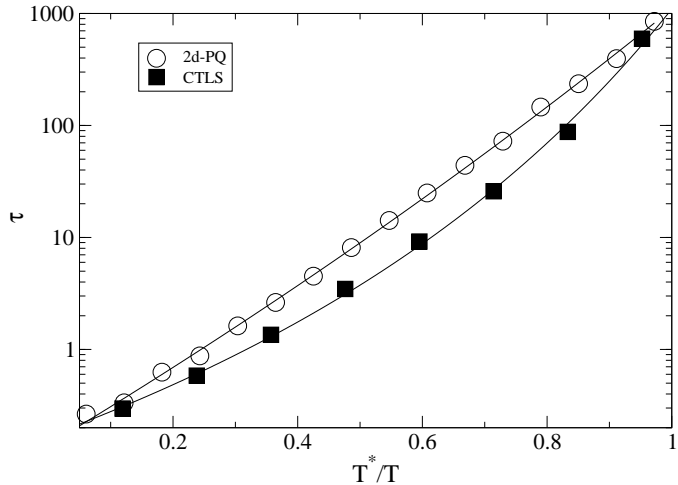}
\vglue 0.2 truecm \caption{Angell plot for the CTLS and the PQ model in $2d$.}
\label{angell}
\end{center}
\end{figure}
This Angell plot \cite{angell} is shown in Fig.~\ref{angell}: it is
clear that the relaxation time of the $2d$-PQ model time is very well
described by an Arrhenius form, $\tau\sim\exp(\Delta/T)$, typical of
strong systems, while the CTLS displays the curved shape of fragile
glasses. To give a quantitative measure of fragility we need to fit
the relaxation time in a Vogel-Fulchner-Tamman form,
\begin{equation}
\tau=\tau_0 \exp\left(\frac{\Delta}{T-T_0}\right)  \ .
\label{vft}
\end{equation}
The fit works pretty well for the CTLS and gives $T_0=0.76$ for
$T\in[1.2:1.7]$, and $T_0=0.90$ for $T\in[1.2:1.4]$.  The quantity
$K_{VFT}=T_0/\Delta$ is a measure of the degree of fragility of the
system \cite{angell}. For the two models we find,
\begin{eqnarray}
2d{\rm PQ:} \qquad K_{VFT}&=& 0.017, \\
{\rm CTLS:} \qquad K_{VFT}&=& 0.095,
\end{eqnarray}
consistent with a high fragility of the CTLS.  In \cite{lipo-bray} it
has been shown that in $d=3$ the PQ model is very well fitted by a
power law, suggesting that it is a fragile model as the CTLS.
Incidentally, we note that an Arrhenius fit of the $2d$-PQ correlation
time gives a barrier size $\Delta=5.8$, at variance with the results
of \cite{juanpe-plaq}, $\Delta=2$. Unfortunately, we were not able to
explain this difference.

\begin{figure}
\begin{center}
\leavevmode
\epsfxsize=3.5 in
\epsffile{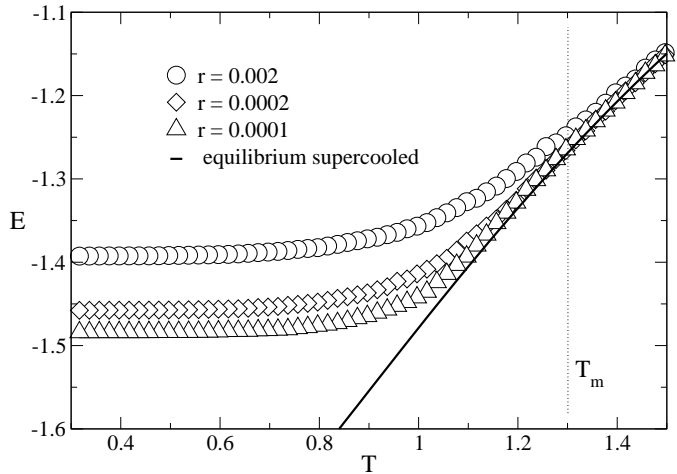}
\vglue 0.2 truecm \caption{Cooling experiments: energy as a function of temperature at
different cooling rates. The cooling protocol is linear,
$T(t)=T_{in}-rt$, from $T_{in}=1.67$ down to $T=0$. Data are shown for
$T\in[0.3,1.5]$ and are stationary at lower temperatures. $L=100$.}
\label{cool}
\end{center}
\end{figure}

From Fig.~\ref{corre} we note that the correlation function $C(t)$
relaxes to zero in a time of order $20\tau$, and this, as we have
seen, is also the order of magnitude of our equilibration time.  Thus,
for $T<1.25$ we are no longer able to equilibrate the system within
our experimental time $t_{\mathrm exp}=10^4$, therefore this is our
glass transition temperature within this time regime.  If, however we
give up the requirement to be at equilibrium, we can of course explore
the low temperature region. For quenches below $T=1.25$ the system
should be considered, according to usual definitions, as an out of
equilibrium glass.  In fact, out-of-equilibrium glasses can be formed
when cooling the liquid at low temperatures at a rate fast enough to
prevent equilibration (the quench from infinite temperature is just the
particular cooling protocol with infinite cooling rate).  In Fig.~\ref{cool} we
show the results of cooling experiments with a linear cooling schedule
$T=T_{in}-rt$ with cooling rate $r=dT/dt$ and $T_{in}=1.67$.  We plot
the system energy as a function of $T$ for different cooling rates. In
the same graph we report the equilibrium supercooled energy
$E_{LQ}(T)$, which of course must be extrapolated below the last
equilibrium temperature. As expected, at high temperatures even for
relatively fast coolings the system stays in equilibrium, whereas
$E(T)$ departs from the equilibrium curve at lower temperatures. The
slower the cooling, the lower the temperature where the energy freezes
in the glassy phase, and this same temperature is the glass transition
$T_g(r)$ corresponding to that particular cooling rate.  As in the
measurement of $\tau$, we have a limit to our experiment set by our
maximum experimental time $t_{\mathrm exp}$: to bring the system through 
the interesting temperature region of width $T\in [0.8:1.3]$, within 
a time $t_{\mathrm exp}$ we can afford at most cooling rates of the order 
of $5 \times 10^{-5}$.

The energy of the configurations reached at $T=0$ upon cooling, as a
function of the cooling rate, is a quantity which is considered as an
important indicator of glassy behaviour. What we find for $r \in
[5\times 10^{-4},0.01]$ is a slow dependence of $E-E_{GS}$ on $r$,
compatible with a power law with low exponent or with logarithmic
behaviour (see Fig.~\ref{ene-gamma} and section V for a wider
discussion of this point), which is very similar to what observed in
real glasses.

Summarizing the results of this subsection, what we observe in the
CTLS for times smaller than $t_{\mathrm \scriptsize sp}\sim 10^4$ MCS
is a phenomenology typical of fragile structural glasses: in the
supercooled liquid phase, $T<T_m$, the correlation function displays
stretched exponential relaxation. The relaxation time $\tau$ grows
very sharply and is well fitted by a power law. Moreover, the exponent
$\beta$ of the stretched exponential fit decreases with decreasing
temperature. A VFT fit of the relaxation time gives a fragility index
much larger than a typical Arrhenius/strong spin system. Finally, on
rapid cooling, the system goes out of equilibrium remaining stuck in
the glassy phase, at a glass transition temperature which is lower the
smaller the cooling rate.

\subsection{Long times dynamics: $t>t_{\mathrm \scriptsize sp}$ and
the metastability limit $\Tsp$}

\begin{figure}
\begin{center}
\leavevmode
\epsfxsize=3.5 in
\epsffile{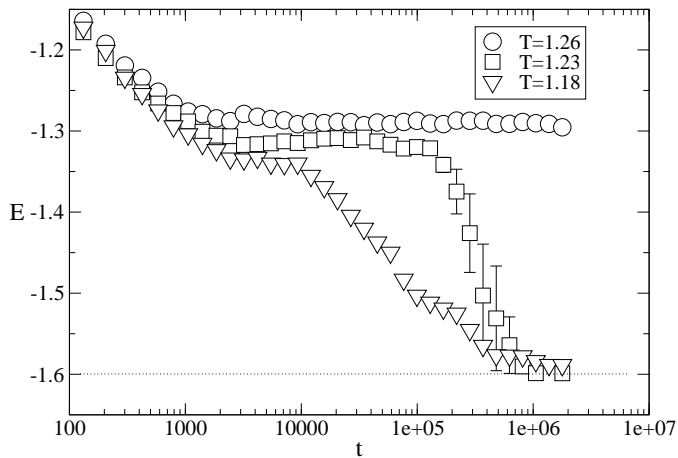}
\vglue 0.2 truecm \caption{Energy per spin close to the spinodal as a function of time. 
The dotted line marks the energy of the  crystal. $L=100$.
Error bars are only shown if larger than symbols size.}
\label{spin-ene}
\end{center}
\end{figure}

We analyze now the behaviour of the system for times longer than
$t_{\mathrm \scriptsize sp}=10^4$.  In Fig.~\ref{spin-ene} we plot the
energy as a function of time for three different temperatures below
$T_m$, with random initial condition.

\begin{figure}
\begin{center}
\leavevmode
\epsfxsize=3.5 in
\epsffile{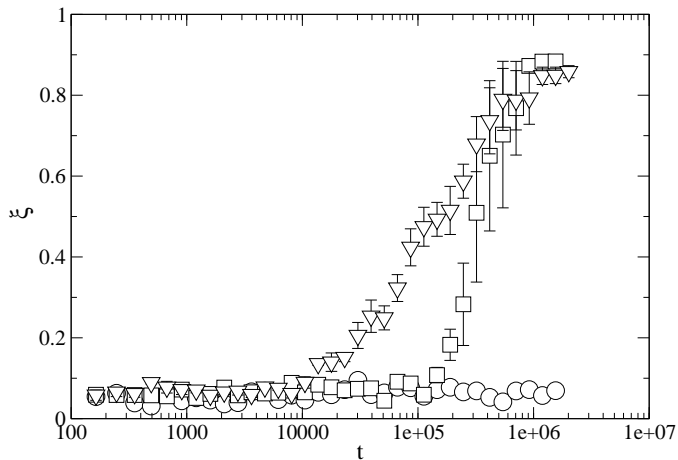}
\vglue 0.2 truecm \caption{Domain size close to the spinodal as a function of time. 
Data and symbols as in Fig.~\ref{spin-ene}.}
\label{spin-xi}
\end{center}
\end{figure}

At the higher temperature, $T=1.26$, the system relaxes in the
supercooled liquid and remains in this phase up to our actual
experimental time, $t_{\mathrm exp}=2\times 10^6$.  On the other hand,
for $T=1.23$, after initial equilibration in the supercooled liquid
phase, the system makes the transition to the crystal.  This fact can
be appreciated in Fig.~\ref{spin-xi}, where we show the behaviour of
the crystal domain size $\xi$, which is clearly highly correlated with
the energy.  Note that at $T=1.23$ the supercooled liquid survives up
to $10^5$ MCS when the transition to the crystal starts, and
crystallization is completed in $10^6$ MCS.  At $T=1.18$, on the other
hand, we see that the supercooled liquid lasts roughly $10^4$ MCS,
while the approach to the crystal takes more than $10^6$ MCS. In other
words, the lower the temperature, the faster is the departure from the
liquid, but the slower is the completion of the crystal.  If we
further decrease the temperature, the plateau corresponding to the
supercooled liquid disappears, and the energy decreases steadily
toward the crystal ground state. The dynamics becomes ever slower in
this temperature regime, and the system remains out of equilibrium up
to very long times (see sect.~\ref{sec-offeq}).

These results can be interpreted in terms of nucleation theory
\cite{nucleation}. Even in the super-cooled liquid phase the system
nucleates droplets of crystal with finite probability.  Most of these
droplets are too small to be stable and soon disappear, however a
critical size exists such that whenever a droplet of that size is
created it will not shrink back, but rather expand and bring the
system into the stable crystal phase.  This phenomenon, critical
nucleation, is characteristic of metastability and determines the
average life-time of the metastable phase. The time needed by the
system to nucleate critical droplets, i.e. the nucleation time
$\tau_{\mathrm nuc}$, depends on temperature and may be very
long. It is often not accessible experimentally, so that the loss of
stability of the metastable phase cannot be observed.  This is not the
case of the CTLS model: for $T=1.23$, $\tau_{\mathrm nuc}$ is indeed
low enough to detect the loss of stability of the super-cooled liquid.
Once critical nucleation takes place, critical crystal droplets start
to grow until all the system has a crystalline structure.
However, the growth of the crystal may become very slow if kinetic constraints
are present.  What we observe in the CTLS is that the lower the
temperature, the faster is crystal nucleation, but the slower is
crystal growth. 

We can formalize the loss of stability of the liquid
by estimating the nucleation time $\tau_{\mathrm nuc}$.  In two
dimensions, the free energy difference due to formation of a crystal
droplet of size $\xi$ in the liquid phase is given by
\cite{nucleation},
\begin{equation}
\Delta(\xi)=A \, \sigma \, \xi - B \, \delta f \, \xi^2,
\end{equation}
where $A$ and $B$ are geometric factors depending on the shape of the
droplet, $\sigma(T)$ is the surface tension, and $\delta
f(T)=F_{LQ}(T)-F_{CR}(T)$ is the bulk free energy difference between
supercooled liquid and crystal below $T_m$. The function $\Delta(\xi)$
has a maximum at a value of the droplet size $\xi^\star$, with an
associated value $\Delta^\star$ of the excess free energy:
\begin{equation}
\xi^\star(T)= \frac{A}{2B} \frac{\sigma(T)}{\delta f(T)}, \qquad
\Delta^\star(T)=\frac{A^2}{4B} \frac{\sigma(T)^2}{\delta f(T)}.
\label{critical}
\end{equation}
For $\xi < \xi^\star$ the system still needs energy to enlarge the
droplet which therefore naturally shrinks its size to zero; for $\xi >
\xi^\star$ on the other hand it becomes favorable for the droplet to
grow.  Therefore, $\xi^\star$ is the critical droplet size and
$\Delta^\star$ the free energy barrier to crystal nucleation.  The
Arrhenius estimate for the time $\tau_{nuc}$ of crystal nucleation is
then
\begin{equation}
 \tau_{nuc}(T)=\tau_0 \exp \left(\frac{A^2}{4B}
 \frac{\sigma(T)^2}{T\,\delta f(T)}\right).
\label{nanu}
\end{equation}

A necessary condition for the existence of the supercooled liquid
phase at temperature $T$ is that the nucleation time $\tau_{\mathrm
nuc}(T)$ is much longer than the time to make an equilibrium
measurement, i.e.\ the equilibration time $\tau_{\mathrm eq}(T)$. In
other words, only if $\tau_{nuc}(T) > \tau_{\mathrm eq}(T)$ can we
speak of a metastable supercooled liquid. We want now to test this
relation in the CTLS. To find $\tau_{\mathrm nuc}$ we need the surface
tension $\sigma$, which we estimate at a reference temperature $T_{ref}$ and assume
that depends weakly on $T$ close to $T_{ref}$. To do this we invert
formula (\ref{nanu}),
\begin{equation}
\sigma^2=4B T_{ref}\,\delta f(T_{ref}) \log[\tau_{\mathrm nuc}(T_{ref})/\tau_0]\,
/A^2 ,
\end{equation}
and after setting $\alpha=T_{ref}\,\delta f(T_{ref}) / T\, \delta f(T)$, we can write
\begin{eqnarray}
\tau_{\mathrm nuc}(T) &\approx& \tau_0 \exp \left\{ \alpha \log \left[
\frac{\tau_{\mathrm nuc}(T_{ref})}{\tau_0} \right]\right\} \\
&=& \tau_{\mathrm nuc}(T_{ref})^\alpha \tau_0^{1-\alpha} \\
&\approx& \tau_{\mathrm nuc}(T_{ref})^\alpha 
\label{nanu1}
\end{eqnarray}
where the last approximation is valid when $T$ is near $T_{ref}$, since
then $\alpha \approx 1$. To have a result as accurate as possible
close to the loss of stability of the liquid, we take $T_{ref}=1.234$,
where $\tau_{\mathrm nuc}=5\times 10^4$. 

Regarding the estimate of $\tau_{\mathrm eq}(T)$, we have already
noticed that the correlation function $C(t)$ drops to zero after
roughly $20$ relaxation times, and this we take as a reasonable
estimate of the equilibration time (it is likely that we are {\it
underestimating} $\tau_{\mathrm eq}$ in this way).  The relaxation
time is given by the power law fit, eq. \ref{power}.

\begin{figure}
\begin{center}
\leavevmode
\epsfxsize=3.5 in
\epsffile{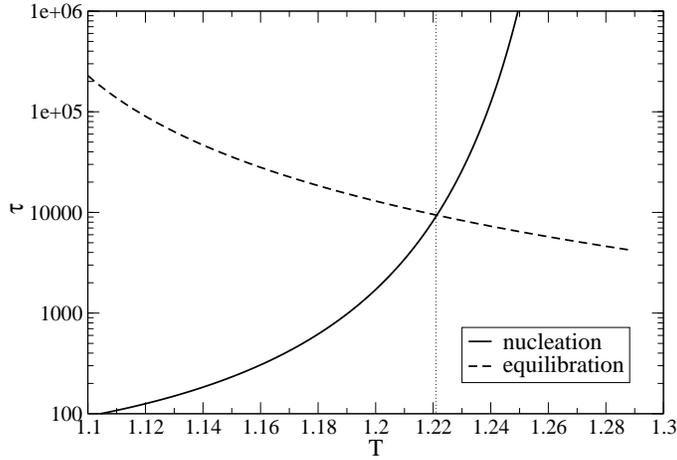}
\vglue 0.2 truecm \caption{Nucleation and equilibration times vs temperature.}
\label{nucleation}
\end{center}
\end{figure}

In Fig.~\ref{nucleation} we plot $\tau_{\mathrm nuc}$ and
$\tau_{\mathrm eq}$ vs.\ $T$ from the above estimates. We see
that the nucleation time drops below the equilibration time at a
temperature $\Tsp=1.22$, which therefore marks the metastability limit
of the supercooled liquid phase. Below $\Tsp$ the liquid can only
be observed on time scales shorter than the equilibration time,
meaning that what we are actually observing is an out-of-equilibrium
glass. Therefore, the equation
\begin{equation}
\tau_{\mathrm nuc}(\Tsp) = \tau_{\mathrm eq}(\Tsp)
\end{equation}
can be considered as a definition of the effective spinodal
temperature $\Tsp$ for the metastable liquid phase. Note that the two
curves cross at a time $t_{\mathrm \scriptsize sp}\sim 10^4$ MCS: this
is the reason why if our maximum experimental time is smaller than
$t_{\mathrm sp}$ we do not detect this metastability limit. As we have
already said, the glass transition temperature $T_g$ is fixed by the
experimental time $t_{\mathrm exp}$ via the relation $\tau_{\mathrm
eq}(T_g)=t_{\mathrm exp}$. Therefore, if $\Tsp < T_g$ the
metastability limit cannot be observed, whereas we detect it if $T_g <
\Tsp$.

\begin{figure}
\begin{center}
\leavevmode
\epsfxsize=3.5 in
\epsffile{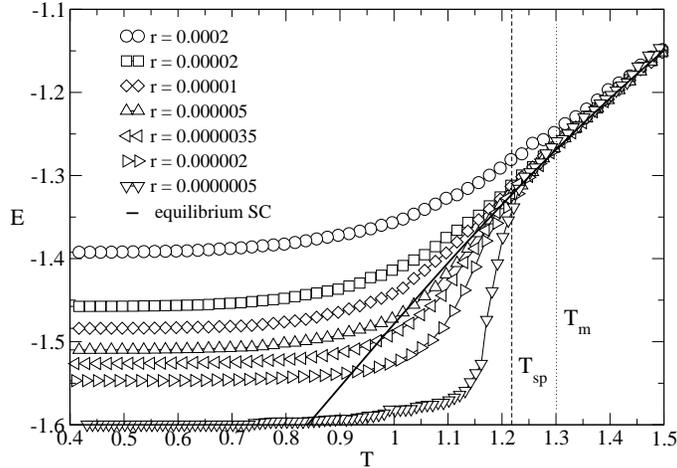}
\vglue 0.2 truecm \caption{Cooling experiments above and below the spinodal cooling
rate $r_{\mathrm sp}$.  Cooling procedure as in Fig.~\ref{cool}. $L=100$.}
\label{cool-total}
\end{center}
\end{figure}

The presence of a lower metastability limit $\Tsp$ is clearly visible
also in cooling experiments. 
If our experimental time exceeds $t_{\mathrm \scriptsize sp}=10^4$ MCS, 
we can afford smaller cooling rates than $r_{\mathrm sp}\sim \Delta 
T /t_{\mathrm \scriptsize sp}$ to explore the relevant region $T\in [0.8:1.3]$.
The results are reported in Fig.~\ref{cool-total}. 
For $r>r_{\mathrm sp}\sim 5 \times 10^{-5}$ the energy departs from
the equilibrium line remaining {\it above} it, while for $r<r_{\mathrm sp}$ 
it drops {\it below} it. We can understand qualitatively 
this behaviour in terms of equilibration and nucleation processes.
For $r>r_{\mathrm sp}$, given the fast 
cooling rate, the system has not the time to equilibrate at low temperatures
where the equilibration time becomes very large and its energy therefore
remains higher than the equilibrium one.
On the contrary, for $r<r_{\mathrm sp}$ the cooling is so slow that 
at low enough temperatures  the system has the time  
to nucleate critical crystal droplets and its energy is  driven 
toward the ground state one, thus becoming smaller than the supercooled 
liquid energy.
In both cases, however, the system eventually remains stuck in an 
off equilibrium phase.

\begin{figure}
\begin{center}
\leavevmode
\epsfxsize=4in
\epsffile{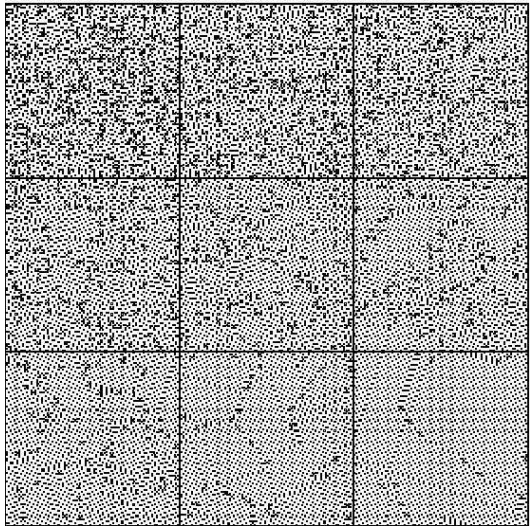}
\vglue 0.4 truecm \caption{Local minima of the CTLS reached at $T=0$ with different cooling 
rates, from $r=2\times 10^{-3}$ (upper left), to $r=2\times 10^{-6}$ (lower right).
$L=100$.}
\label{conf}
\end{center}
\end{figure}

\begin{figure}
\begin{center}
\leavevmode
\epsfxsize= 3.5in
\epsffile{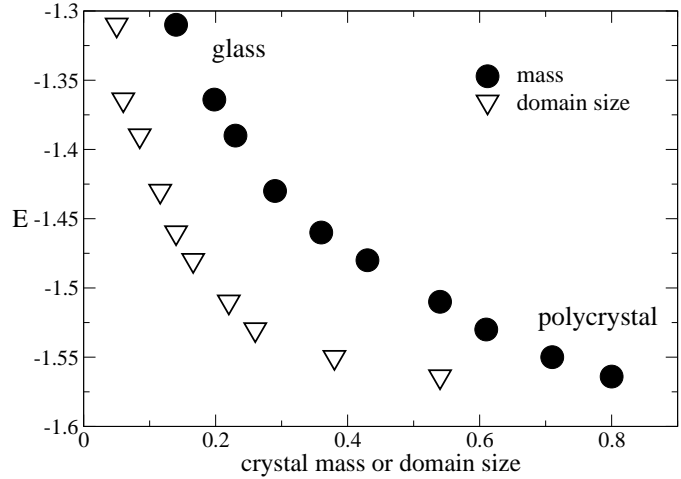}
\vglue 0.2 truecm \caption{Energy of the configurations reached at $T=0$ during cooling
experiments as a function of the mass and the domain size. $r\in
[10^{-6},10^{-3}]$, $L=100$.}
\label{cool-mass-xi}
\end{center}
\end{figure}

This experiment poses an interesting question about the actual nature
of the out-of-equilibrium glassy phase. 
For fast cooling rates the system remains stuck at a high energy level, 
into a clearly disordered, or glassy-like, phase (Fig.~\ref{conf}, upper 
left). On the other hand, for the slowest cooling rates
the system approaches the ordered crystal phase, even though it 
remains stuck in an out-of-equilibrium polycrystalline configuration,
as Fig.~\ref{conf} (lower right) shows.  
The decrease in energy, and the associated increase in crystalline order
that we observe as we slow down the cooling is a continuous process,
making it very difficult to distinguish what is a glass and what is a
highly disordered polycrystal: every possible structural way to detect 
crystalline order in the upper left configurations of Fig.~\ref{conf} would 
give the same result as for the liquid phase. Note, finally,  that all the 
configurations in Fig.~\ref{conf} are stable local minima of the Hamiltonian 
reached at $T=0$.
This {\it structural continuity} from glass to polycrystal can be appreciated 
more quantitatively in Fig.~\ref{cool-mass-xi}, where we plot the energy of the asymptotic
configurations the system reaches at $T=0$, as a function of the
corresponding crystalline domain size $\xi$ and crystal mass $m$, for
various cooling rates $r$.  All the points on this plot corresponds to
configurations obtained with a well-defined cooling
experiment. Moreover, all these configurations are (local) minima of
the energy. As we can see, the data interpolate without gaps between
crystalline (or polycrystalline) configurations with low $E$ and high
$\xi$ or $m$, to glassy configurations with high $E$ and low $\xi$ or
$m$.  In this sense we may say that in the present system there is no
qualitative difference between the off-equilibrium disordered, i.e.\
glassy, regime, and the crystal growth regime. On the other hand,
thanks to the presence of the pseudo-spinodal transition, there is a
clear distinction between {\it equilibrium} liquid and crystal phases.

\subsection{Metastability limit and relaxation time divergence:
$\Tsp$ vs $T_c$}

It has been shown in \cite{lipo-bray} that in the PQ model the
temperature $T_c$ where a power law fit of the relaxation time locates
a divergence, is very close to the metastability limit $\Tsp$. The
interpretation of this fact given in \cite{lipo-bray} is very clear:
in mean-field the PQ model has a sharp spinodal temperature at the
point where the liquid metastable state disappears, and this
transition is accompanied by a divergence of the correlation
times. Therefore, in \cite{lipo-bray} it is argued that in the
finite-dimensional version of the PQ model the temperature $T_c$ is
actually a remnant of the mean-field spinodal. The same view has been
advanced for off-lattice models: in \cite{klein2d} the behaviour of a
supercooled two-component Lennard Jones system in $d=2$ has been
interpreted as due to the influence of a pseudospinodal, while in
\cite{klein3d} arguments are given to explain why in $d=3$ spinodal
effects may be much harder to detect.

The situation is very similar in the CTLS, even though our definition
of $\Tsp$ makes it indispensable to distinguish between the two
temperatures.  The temperature $T_c$ where the relaxation time would
diverge is not actually a true critical temperature: when the
temperature is decreased the equilibration time increases and finally
becomes as large as the nucleation time at $\Tsp$, when $\tau_{\mathrm
eq}=\tau_{\mathrm nuc}$ is still finite. Therefore we have by
definition $\Tsp>T_c$.  This is simply due to the finite-dimensional
nature of the CTLS. In mean-field systems a metastable state has
infinite life-time and is separated from the stable phase by an
infinite free energy barrier: the spinodal temperature, which marks
the disappearance of the metastable state, corresponds to the
temperature where the barrier goes to zero, and is naturally
associated with a divergent time-scale. In a finite dimensional system
like the CTLS, on the contrary, the metastable phase (the liquid) is
separated from the stable one (the crystal) by a finite free energy
barrier ---see eq.~(\ref{critical})--- and always has a finite
lifetime.  An effective spinodal temperature can only be defined by
comparing the different and finite time-scales present in the system,
as done in the previous subsection. However, the physical
interpretation of \cite{lipo-bray} may remain valid, i.e.\ the
increase of the relaxation time close to $\Tsp$ may be an effect of
approaching incoming instability of the supercooled liquid.

We stress again that had our maximum experimental time been shorter
than $t_{\mathrm \scriptsize sp}=10^4$ MCS, we would not have detected
the metastability limit at $\Tsp$, while we still would fit the
equilibrium relaxation time as a power law with critical temperature
$T_c$. In that case, it would have been harder to realize the
connection between divergence of the relaxation time and loss of
stability.

\subsection{Kauzmann's resolution of the Kauzmann paradox}

In 1948, W.~Kauzmann outlined in a famous paper \cite{kauzmann} the
paradox now named after him. In the supercooled phase the viscosity of
the system increases with decreasing temperature, until at the glass
transition $T_g$ relaxation time becomes too long and equilibration
cannot be achieved within experimentally accessible times. However,
Kauzmann noted that if the entropy of a supercooled liquid is
extrapolated below $T_g$, it becomes equal to the crystal entropy at a
temperature $T_s>0$, and even smaller than zero if extrapolated
further. This {\it entropy crisis} is never actually observed, because
the glass transition intervenes before. However, Kauzmann found it
paradoxical that it was just a kinetic phenomenon (the glass
transition) that saved the liquid from a thermodynamic nonsense.

In the context of the Adams-Gibbs-DiMarzio theory \cite{adams}, the
entropy crisis has however an interpretation: the entropy difference
between crystal and liquid is related to the configurational entropy
$\Sigma$, that is the entropic contribution due to the presence of an
exponentially high number of different glassy minima. The vanishing of
$\Sigma$ at $T_s$ signals a thermodynamic transition to a new phase,
characterized by a sub-exponential number of glassy states, separated
by infinite free-energy barriers. This picture is exact for some
mean-field spin-glass systems \cite{pspin}, and it may be the correct
resolution of the Kauzmann paradox even for real structural glasses.
According to Adam-Gibbs-DiMarzio, the entropy crisis is accompanied by
a divergence of the relaxation time at $T_s$ given by $\tau \sim
\exp(A/T \Sigma)$.  Expanding linearly the configurational entropy
close to $T_s$ a Vogel-Fulchner-Tamman behaviour is obtained (see
eq.~\ref{vft}) with $T_0=T_s$. $T_0$ and $T_s$ have been observed to
be quite close in various systems \cite{vetri}, a fact that has been
advocated as an indirect argument in favour of this interpretation.

Despite recent analytic and numerical work supporting the entropy
crisis scenario \cite{giorgio-marc,sciortino-entropy}, there is
another way to avoid the Kauzmann paradox, which, interestingly
enough, was proposed by Kauzmann himself \cite{kauzmann}.  He rejected
the idea of a thermodynamic glassy phase, and of a transition at
$T_s$. What Kauzmann hypothesized is the existence of a metastability
limit of the supercooled liquid phase, below which crystal nucleation
becomes faster than liquid equilibration.  More specifically, he defined
an effective spi\-no\-dal temperature $T_{\rm sp}>T_s$ below which
{\it ``the free energy barrier to crystal nucleation becomes reduced
to the same height as the barrier to simpler motions''}.  Below
$T_{\rm sp}$ the supercooled liquid is operationally meaningless and
thus the paradox is avoided.

What we have shown in this paper is that the CTLS is a
system with a super-cooled phase and a fragile-glass phenomenology
that precisely reproduces the scenario hypothesized by Kauzmann.  If we look at
the behaviour of the extrapolated excess entropy $S_{LQ}-S_{CR}$ we
find that in this model $T_s$ and $T_0$ are actually very similar (see
sections III and IV-A).  However, despite this fact, no entropy crisis
does actually take place.  The metastable liquid ceases to exist below
$\Tsp=1.22$ and thus extrapolations of thermodynamic quantities in the
supercooled phase below $\Tsp$ are meaningless.  Therefore, the
paradox is avoided: nucleation takes over when the relaxation time
becomes larger than the nucleation time.  Of course, we can still
speak of an out of equilibrium glass below $\Tsp$, which can be
obtained if we cool fast enough our system.

\begin{figure}
\begin{center}
\leavevmode
\epsfxsize=4in
\epsffile{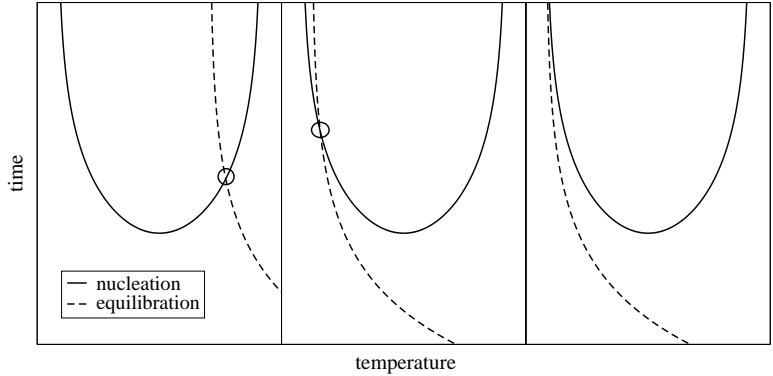}
\vglue 0.2 truecm \caption{Pictorial view of three different scenarios: 
Left: $\tau_{nuc}$ and $\tau_{eq}$
cross at a spinodal temperature $\Tsp$, which is larger than the 
temperature where $\tau_{\mathrm nuc}$ has a minimum, as in the CTLS.
Centre: $\tau_{nuc}$ and $\tau_{eq}$ cross at a spinodal temperature 
$\Tsp$, which is smaller than the temperature where $\tau_{\mathrm nuc}$ 
has a minimum. Right: $\tau_{nuc}$ and $\tau_{eq}$ both diverge at the 
same temperature $T_s=\Tsp$. The circles indicate the spinodal temperature
$\Tsp$.} 
\label{nucl2}
\end{center}
\end{figure}

As we have already stressed, the metastability limit may prove
impossible to observe experimentally if the equilibration time at
$T_{\rm sp}$ is much larger than the experimental time, that is if
$T_{\rm sp}\ll T_g$. One is then tempted to suggest that the Kauzmann's
resolution of the Kauzmann paradox that we have explicitly analyzed in
the CTLS model possibly applies to many other systems, but is actually
out of experimental recognition.  However, we must be careful in
judging how general this behaviour can be. The rest of this Section is
devoted to a discussion of this point. 
\begin{figure}
\begin{center}
\leavevmode
\epsfxsize=3 in
\epsffile{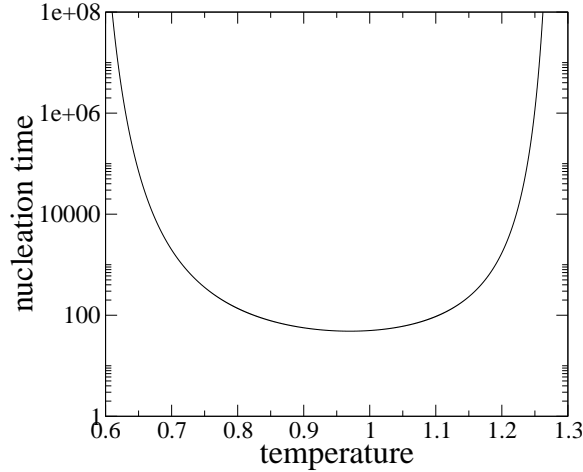}
\vglue 0.2 truecm \caption{Extrapolated nucleation time of the CTLS model vs.\
temperature. A minimum is clearly visible at $T=0.98$.}
\label{nucl1}
\end{center}
\end{figure}   
In Fig.~\ref{nucl1} we see that the 
nucleation time $\tau_{\mathrm nuc}$ of the CTLS has a
minimum at a temperature $T=0.98$, below which it starts increasing
again. This is simply due to the balance between the free energy
barrier $\Delta^\star(T)$ and $T$ in the Arrhenius factor, and it is
certainly a very model-dependent property.  
\footnote{We have assumed a
constant surface tension close to $T_{ref}=1.23$, and thus we cannot trust
too much $\tau_{\mathrm nuc}$ far from that point. Moreover, we have
ignored any possible temperature dependence in the prefactor of
Arrhenius's law (\ref{nanu}).}
In the CTLS all this is of little relevance, because $\tau_{\mathrm nuc}$ 
and $\tau_{\mathrm eq}$ cross at a much higher temperature, as shown in 
Fig.~\ref{nucleation} and also in Fig~\ref{nucl2}-Left.
Yet, in general the interplay between $\sigma(T)$, $\delta f(T)$
and $T$ may be such that a minimum of $\tau_{\mathrm nuc}(T)$ actually
occurs {\it and} that at this minimum $\tau_{\mathrm nuc} \gg \tau_{\mathrm
eq}$.  However, even in this case it may still happen that the curves cross
below the minimum of $\tau_{\mathrm nuc}$, as shown in Fig.~\ref{nucl2}-Centre. 
In this case the Kauzmann paradox would still be resolved by the
metastability limit, exactly as in CTLS. Note that this particular scenario
may actually prove to be quite deceiving: at temperatures smaller than the 
minimum of the nucleation curve one can think that the `dangerous' zone
for crystal nucleation has been safely crossed and that crystallization
is therefore ruled out at lower temperatures. When $T_g$ is met at lower 
temperatures, one would need some very careful extrapolations of both
$\tau_{\mathrm nuc}$ and $\tau_{\mathrm eq}$ to recognize that a 
metastability limit actually exists at $\Tsp \ll T_g$.

But is it possible that $\tau_{\mathrm nuc} \gg \tau_{\mathrm eq}$ at
{\it all} temperatures?  The answer is certainly yes for strong liquids,
where $T_s=0$, and thus $\tau_{\mathrm eq}$ stays finite for $T>0$;
but this case is of little interest for us since then the Kauzmann
paradox does not arise. In fragile liquids, however, where $\tau_{\mathrm eq}$
is supposed to diverge at $T_s$, the condition for the existence of the liquid, 
$\tau_{\mathrm nuc} \gg \tau_{\mathrm eq}$, requires
that $\tau_{\mathrm nuc}$ diverges at a temperature $T_{nuc}\ge T_s$. 
In our opinion it is possible that $T_{nuc}=T_s$, since at low temperatures 
the contribution of $\tau_0$ in eq.~\ref{nanu} can be important due to kinetic 
constraints, as shown in \cite{turnbull}. In nucleation theory, this prefactor is 
taken as proportional to the viscosity, in which case $\tau_{\mathrm nuc}$ would
diverge when the viscosity does, that is at $T_s$. 
But it is hard to imagine how $\tau_{\mathrm nuc}$
could diverge {\it strictly above} the thermodynamic transition temperature $T_s$. 
Therefore, it seems to us that the best one can 
have is $\Tsp = T_s$, as in Fig.~\ref{nucl2}-Right. 

Let us stress once again the fact that
crystal nucleation is a different process from crystal growth: saying
that nucleation is fast, is not the same as saying that the critical
crystal droplets will grow quickly, since the speed of growth will
strongly depend on kinetic considerations.  From an experimental point
of view, it is clear that having a very long nucleation time (longer
than the experimental time) and fast growth, is not much different
from having a fast nucleation of very small, undetectable, crystal
droplets, which however grow exceedingly slowly. 
In order to avoid crystallization and form the glass we have either to
avoid nucleation, or to ensure that crystal growth is kinetically
blocked.  To clarify this point, we can ask how long is the
time $t_x$ the system takes to develop a substantial amount of
crystalline order, large enough to prove that crystallization has
started.  We can answer this question by following the time evolution
of the crystal mass $m(t)$ at various temperatures, and defining $t_x$
from the relation $m(t_x)\geq m_{th}$, where $m_{th}$ is an arbitrary
threshold.  The function $t_x(T)$ is normally called in the context of
supercooled liquids Time-Temperature-Transformation (TTT) curve, and
it is the only experimentally available means to check how likely is
crystal formation at a given temperature. Note that the value of
$t_x(T)$ will strongly depend on $m_{th}$, and it cannot therefore be
{\it quantitatively} compared to $\tau_{eq}$.

\begin{figure}
\begin{center}
\leavevmode 
\epsfxsize=3.5in 
\epsffile{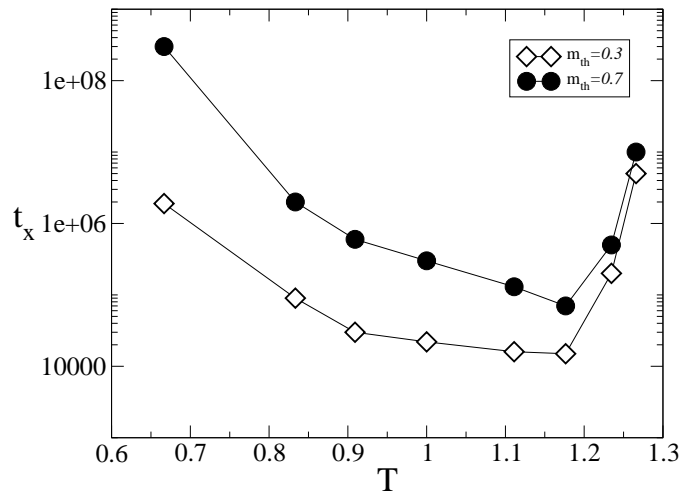}
\vglue 0.2 truecm \caption{TTT curves for two different values of the threshold value
$m_{th}$.  L=100.}
\label{ttt}
\end{center}
\end{figure}
In Fig.~\ref{ttt} we plot the TTT curves of the CTLS for two different
values of $m_{th}$. As in realistic supercooled liquids, the TTT
curves have a minimum at a temperature $T_x$: here crystallization has
the fastest rate, in the sense that at $T_x$ we have to wait the
minimum possible time to reach a level $m_{th}$ of
crystallization. The origin of this minimum is obvious in CTLS: for
high temperatures $t_x$ is large because {\it crystal nucleation} is
very slow, i.e. because $\tau_{nuc}$ is very large, whereas for small
temperatures $t_x$ is large because {\it crystal growth} is very slow,
which is essentially due to the high viscosity of the system (i.e. to
the same reason why, if the liquid still existed, $\tau_{\mathrm eq}$
would be very large).  Thus, in CTLS the two branches of the
TTT curve correspond to two different phenomena, that is nucleation
and growth. Not surprisingly we find $T_x\sim \Tsp$.  
On the other hand, in a system where the scenario depicted in Fig.~\ref{nucl2}-Centre 
holds, also the left branch of the TTT curve for $T<T_x$ would be due to slow
nucleation \cite{kusalik,sciortino}, until $\Tsp<T_x$ is reached.
In this case the minimum of the TTT curve would be an effect of the 
minimum of $\tau_{\mathrm nuc}(T)$ described above. Below $\Tsp$, however, 
the TTT curve would again increase due to slow crystal growth.

It may be argued that there is little experimental
difference between the case where crystallization at low $T$ is
blocked because of arrested nucleation or arrested growth.  However,
from a theoretical point of view these two situations are quite
different: in the first case the equilibrium liquid phase exists, at
least on certain time scales, whereas in the second case it does not,
the only liquid phase is the out-of-equilibrium glass and the Kauzmann
paradox is avoided. We can characterize these two scenarios as {\it
equilibrium arrested nucleation} vs {\it off-equilibrium arrested
growth}.

\section{Off-equilibrium behaviour: domain growth and aging}

\label{sec-offeq}

In this section we study the dynamical behaviour of the CTLS after a
quench to $T<\Tsp$. We shall first focus on one-time quantities, and in
particular on the behaviour of energy, crystal domain size and crystal mass. 
Then we will analyze two-time quantities to understand whether our system 
exhibits aging, and of what kind. The key result of this section is that 
there are two time regions for off-equilibrium dynamics. In the earlier time 
regime there is the formation of a large number of critical crystal droplets 
all over the system. As we have seen, this mixture of crystallites cannot be
structurally distinguished from a glass or liquid configuration, as long as
the crystallites are very small. Their growth is controlled by activation, and
thus this regime will last longer the lower the temperature. In this phase, 
which we call {\it bubbling}, off-equilibrium dynamics is similar to what 
found in structural glasses. The later time regime kicks in when the energy 
barriers to crystal growth have been crossed, so that crystal domains are 
large enough to be structurally recognizable. This phase we call {\it coarsening}, 
and off-equilibrium dynamics in this regime is consistent with models where 
simple coarsening occurs, as the Ising model. 
Not surprisingly, the minimum time we have to wait to see the coarsening regime 
in our system (and thus realize that crystal domains are actually growing) 
is as long as the time needed to directly see the loss of stability of 
the liquid, that is $t_{\mathrm sp}$.
In other words, the dynamical checks for the growth of the crystal are
as inefficient as the structural ones we considered in the former Section.

\subsection{One time quantities}

\begin{figure}
\begin{center}
\leavevmode
\epsfxsize=3.5in
\epsffile{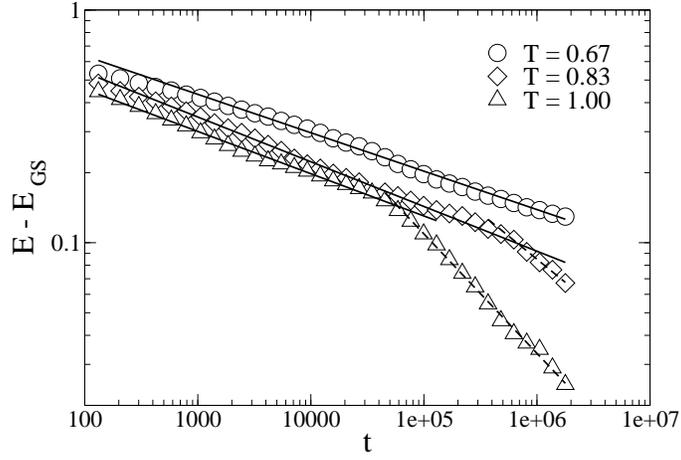}
\vglue 0.2 truecm \caption{Excess energy as a function of time after a quench at $T<\Tsp$. 
Full lines: power law fit of the early (slow) time region:
$\nu = 0.17$ for $T=0.67$; $\nu = 0.19$ for $T=0.83$; $\nu = 0.18$ for $T=1.00$.
Dashed lines: power law fit of the late (fast) time region:
$\nu = 0.40$ for $T=0.83$; $\nu = 0.51$ for $T=1.00$. $L=100$.}
\label{fit-ene}
\end{center}
\end{figure}

As seen in the previous sections, when quenched to temperatures in the
range $\Tsp<\ T<\ T_m$, the system relaxes into the supercooled
liquid phase and here remains up to times of order $\tau_{nuc}$ (see
Fig.~\ref{meta-ene}). On the other hand, below the spinodal temperature 
$\Tsp$ the supercooled liquid no longer exists and after a quench the energy 
very slowly evolves toward its crystalline ground state value. The rate 
of decay of the energy gives a useful characterization of the 
off-equilibrium dynamics. In Fig.~\ref{fit-ene} we report the behaviour 
of the excess energy $\delta E=E-E_{GS}$ as a function of time, together 
with power law fits of the form $\delta E=A\, t^{-\nu}$.  For all temperatures
$T<\Tsp$ we find an initial time regime where $\nu < 0.2$, which is compatible
with what found in some model systems of structural glasses (see for
ex. \cite{kob-barrat,andrej}). This first time regime is followed by a
second late time regime where the energy decreases faster, with $\nu \sim 0.5$. 
However, the data show that the crossover to this second time regime 
takes place later the lower the temperature. At $T=0.67$ we are unable to
enter this late time region within our maximum experimental time, and we 
only observe the slow regime.

The existence of these two different regimes can be related 
to the presence of {\it activated coarsening}, as is commonly
found in frustrated lattice spin models.
\cite{sethna,lipo}.  Below the spinodal temperature the system 
rapidly nucleates small stable crystal domains, which tend to grow.
However, due to the high viscosity, local spin rearrangements are 
difficult, and moving a domain boundary may be rather costly. Thus, 
contrary to the case of standard coarsening in the Ising model, when 
frustration is present domain boundaries are pinned, and domain growth 
requires overcoming some energy barriers. The system therefore needs 
activation to grow crystal domains, and this may lead to a non-standard
decay of the energy as a function of time.

More precisely, according to \cite{mazenko} we must distinguish two 
classes of activated coarsening, depending on whether or not the maximum 
barrier $\Delta$ to domain growth depends on the linear domain size 
$L\xi(t)$, and therefore on $t$ itself. In the first class of systems  
$\Delta$ does not depend on $\xi(t)$. In this case the system evolves 
slowly within a time scale $t_\Delta\sim\exp (\beta \Delta)$, 
while for $\log t > \log t_\Delta$ barriers are no longer a problem, 
and a standard power law behaviour is recovered,
i.e. $E-E_{GS}\sim t^{-1/2}$ (we are only considering systems with 
nonconserved order parameter).
On the contrary, if the barrier to domain growth is proportional to $\xi(t)$, 
then it is larger the longer the time. In this second class of activated 
coarsening one expects a logarithmic asymptotic dependence of the  
energy on time.

\begin{figure}
\begin{center}
\leavevmode
\epsfxsize=3.5in
\epsffile{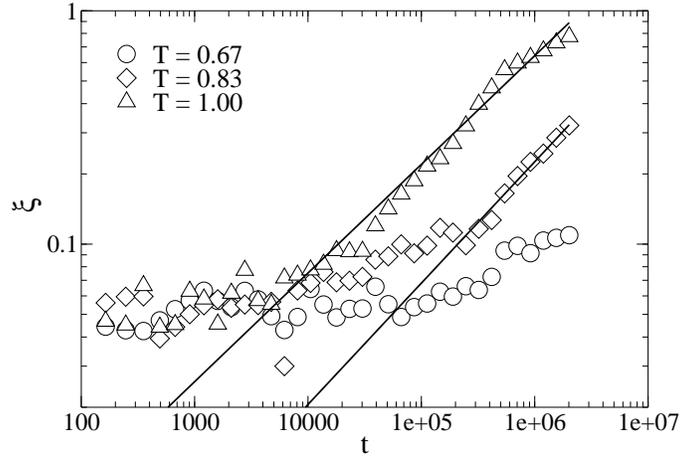}
\vglue 0.2 truecm \caption{Domain size as a function of time after a quench at $T<\Tsp$. 
Full lines correspond to power law fits of the later fast regime, 
$\xi \sim t^{\nu}$. The exponents 
are $\nu\sim 0.45$ for $T=1.0$ and $\nu\sim 0.52$ for $T=0.83$.  $L=100$.}
\label{fit-domain}
\end{center}
\end{figure}

\begin{figure}
\begin{center}
\leavevmode
\epsfxsize=3.5in
\epsffile{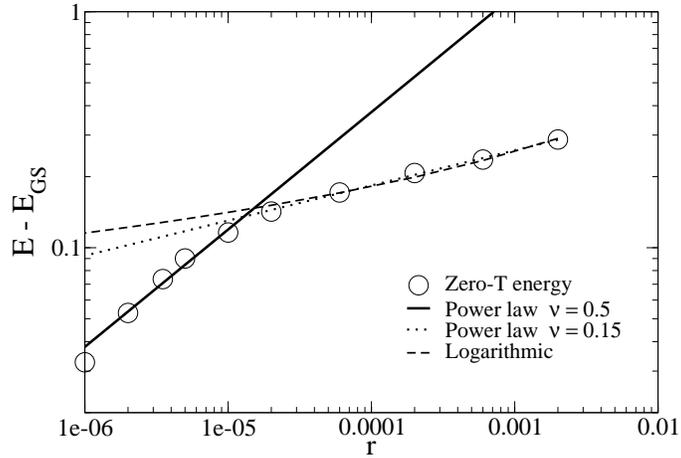}
\vglue 0.2 truecm \caption{ Energy of the configurations reached at $T=0$ during cooling experiments
 (see Fig.~\ref{cool-total}) 
as a function of the cooling rate, together with power law and logarithmic fits. $L=100$.}
\label{ene-gamma}
\end{center}
\end{figure}

The data regarding the energy in Fig.~\ref{fit-ene} show
that, after an early slow regime, the standard $t^{-1/2}$ decay of the energy 
is recovered. This fact suggests that our model belongs to the first 
class of activated coarsening, and that the early time regime is indeed
connected by activation to a constant energy barrier to domain growth
\footnote{We note that in the models studied in \cite{sethna,lipo}, a
detailed analysis of the barrier to growth showed that 
these two models actually belong to the second class of activated coarsening.  
}. 
A similar conclusion can be inferred from the behaviour of $\xi(t)$.  In
figure ~\ref{fit-domain} we show for three temperatures below $\Tsp$
the behaviour of $\xi$ as a function of time. Also in this case for the higher
temperatures we find quite clearly two different regimes: a first
one in which $\xi$ does not increase, but rather fluctuates around its effective
zero, and a second one where it increases quite sharply.  The statistics is
not sufficient to envisage a clear asymptotic dependence, however a
power law fit gives for the late time regime an exponent close to
$1/2$.

The behaviour of the system when we perform a cooling experiment 
is consistent with the interpretation above.  
The configurations reached at $T=0$ for low cooling
rates exhibit the typical pattern of a coarsening system
(Fig.~\ref{conf}) and are stable, indicating that coarsening is
indeed activated. Besides, if we plot the excess energy $\delta E$ 
as a function of the cooling rate $r$, once again we find two regimes
(fig. ~\ref{ene-gamma}): for fast coolings (short time) we
have a slow energy decay, which can be fitted either with a
logarithm, or with a power law $\delta E\sim r^\nu$, with 
$\nu \sim 0.15$. On the other hand, for slow 
coolings (long time) we find $\nu\sim 1/2$, as in standard 
coarsening for the Ising model. This behaviour of $\delta E(r)$ 
is what we expect for activated coarsening of the first class 
and can be qualitatively explained in the following way.
As we have seen, below $T_{\mathrm sp}$ the system exhibits a non trivial dynamical 
behaviour where the energy evolution is determined by the formation and growth of pinned crystal 
domains. In a cooling experiment it is then crucial how much time  
the system spends below the spinodal temperature before remaining trapped in the 
asymptotic state. If we call $T_f$  the temperature where the system reaches the asymptotic 
$T=0$ energy value 
\footnote{
$T_f$ can be estimated as the temperature where thermal activation becomes extremely rare even
for simple rearrangements, i.e. $\delta/T_f \gg 1$  where $\delta$ is the 
{\it smallest} barrier to kinetic motion.
},
then at a given cooling rate $r$ the time elapsed from  $T_{\mathrm sp}$ to $T_f$ is
given by $t(r) = (T_f-T_{\mathrm sp}) / r$.  During this time the energy of the system 
evolves 
approximately as described previously in this section, i.e. $\delta E=A t^{-\nu}$.  
Therefore we have  $ E(r)- E_{GS} \sim t(r)^{-\nu}\sim r^\nu$ and the two regimes observed 
in Fig. ~\ref{ene-gamma} are nothing else than a manifestation of the two regimes already 
discussed  for   Fig. ~\ref{fit-ene}.
The exponent $\nu$ corresponds to the first regime value  if 
$t < t_{\mathrm sp}$, that is if $r >  r_{\mathrm sp}$, while for $r <  r_{\mathrm sp}$ the 
standard value $\nu=1/2$ is recovered.

\subsection{The bubbling and coarsening regimes}

The two regimes we have presented above can be explicitly related 
to different patterns of domain dynamics. In order to do this we
compare the different time dependence of the domain size $\xi$ and
crystal mass $m$ \footnote{We recall that both these quantities are
define as densities. The actual linear length of a domain is $L\xi$,
and the total Crystal mass is $L^2m$.}.
Up to now we used both quantities as equivalent markers of crystal 
order, but in fact in the growth phase their role is quite different.

\begin{figure}
\begin{center}
\leavevmode
\epsfxsize=3.5in
\epsffile{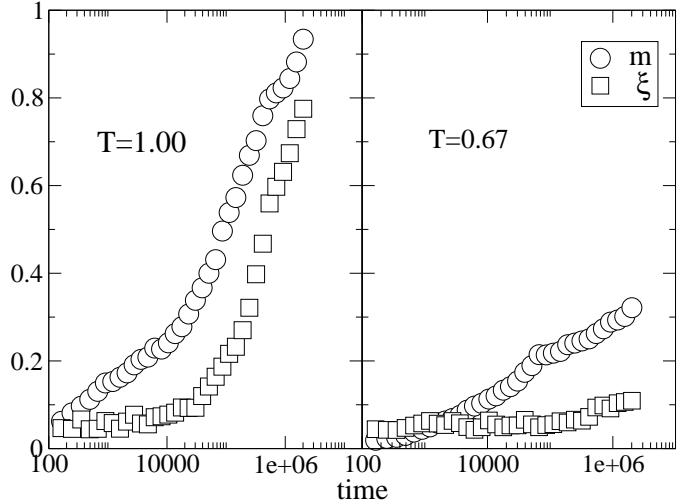}
\vglue 0.2 truecm \caption{Mass and domain size as a function of time for two different 
temperatures. $L=100$}
\label{mass-xi}
\end{center}
\end{figure}

After a quench to $T\ll \Tsp$ we expect that fast
nucleation will lead to an increase in $m$ due to the formation of
many crystallites, but that the typical domain size $\xi$ will remain
very small, actually smaller than our resolution. This picture
is confirmed by Fig.~\ref{mass-xi}, where we plot $\xi$ and $m$ as a
function of time for two different temperatures below $\Tsp$. In both
cases we see that the initial increase of $m$ is faster than that of
$\xi$. For $T=1.0$, which is quite close to $\Tsp=1.22$, we can see a
rise of $\xi$ only for times longer than $10^5$ MCS, while for
$T=0.67$ there is practically no increase of $\xi$ up to $t=10^6$ MCS,
while a clearly detectable increase of $m$ can be seen.

\begin{figure}
\begin{center}
\leavevmode
\epsfxsize=3.5in
\epsffile{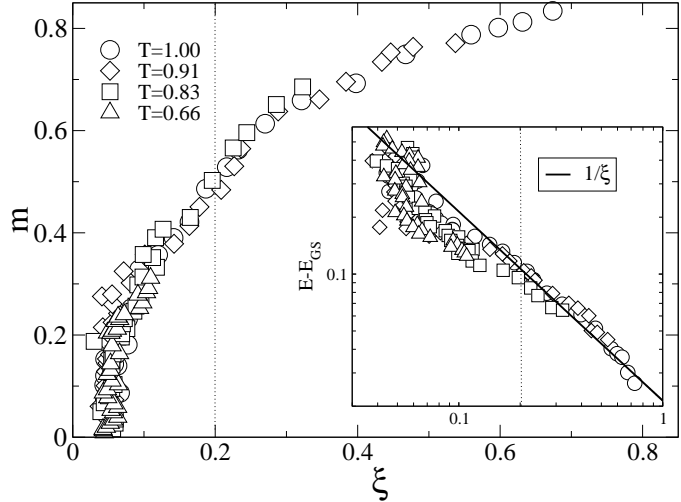}
\vglue 0.2 truecm \caption{Crystal mass as a function of the domain size parametrically in $t$, 
after a quench at various temperatures. 
Inset: excess energy as a function of the domain size. Full line:
$1/\xi$ fit. All runs are $2 \times 10^{6}$ MCS long, $L=100$. }
\label{domain-m-xi}
\end{center}
\end{figure}

To further support this picture we plot in Fig.~\ref{domain-m-xi} the mass 
$m$ as a function of domain size $\xi$ 
for many different temperatures below $\Tsp$, and for different times.  
All these data collapse on a master curve, which shows that for short times
(small $\xi$ and $m$) the crystal mass grows much faster than the
domain size, whereas for long times $m$ tends to saturate, and $\xi$
start increasing rapidly.

We can therefore define two time regimes for crystal domain growth. 
For shorter times, in what we may call a {\it bubbling} regime, 
many small crystallites form at the expenses of a
'liquid' background, their linear size growing very little with time, 
whereas the total crystal mass grows quite sharply \cite{glacial}.
In this phase domains are so small that they are not even well defined, 
in the sense that their linear size $L \xi$ is of the same order as the width
$a$ of the interfaces (or domain walls) among them. In this regime 
the excess energy is dominated by the regions not yet crystallized, and the
interfacial and domain contributions are of the same order.  For
longer times, the largest part of the system is crystallized, so the 
mass $m$ saturates,
and the small domains start growing one at the expenses of each
other. 

\noindent
Therefore a proper {\it coarsening} regime starts with a clearly
increasing $\xi$, and $L\xi \gg a$.  Now domains are well defined 
and the excess energy is concentrated in the interfaces. These are 
now much thinner than the domain size and can be considered as lower 
dimensional manifolds. This must give,
\begin{equation} E(t)-E_{GS} \sim
\frac{\xi(t)^{d-1}}{\xi(t)^d} = \frac{1}{\xi(t)} \ , 
\end{equation}
as in standard coarsening\cite{bray,mazenko}.
In Fig.~\ref{domain-m-xi} we plot the excess energy $E(t)-E_{GS}$
vs.\ $\xi$ for a number of temperatures and times. Consistently with 
our conclusions, for shorter times there is a drop in energy with
basically no increase in $\xi$. On the other hand, for $\xi>0.2$ the
$1/\xi$ law fits reasonably well the data, suggesting that this is the
late-time coarsening regime. Note that $\xi>0.2$ implies $m>0.5$ (see
Fig.~\ref{domain-m-xi}).  We could thus schematically say that the
coarsening regime sets in when the total crystal mass exceeds $50\%$.
This definition of the two regimes is consistent with the behaviour
displayed by $E(t)$: for $T=1$ the system enters the coarsening regime
(i.e. $m>0.5$ and $\xi>0.2$) for $t \sim 40000$ (see
Fig.~\ref{mass-xi}), precisely on the same time scale the energy
(Fig.~\ref{fit-ene}) passes from a power law behaviour with low
exponent to a faster $t^{-1/2}$ behaviour. For $T=0.67$ on the other hand
$\xi<0.2$ for our longest time, and correspondingly the energy always
obeys a slow power law decay.

\subsection{Two time quantities}

We now focus on the behaviour of two-time quantities, for quenches
below $T_m$.  In equilibrium dynamics, correlation functions
$C(t,t_w)=\langle A(t_w) A(t)\rangle$ and their associated response
functions $R(t,t_w)= \delta \langle A(t) \rangle / \delta h(t_w)$,
($A$ is a generic variable and $h$ is the field conjugate to it)
depend only on the difference $t-t_w$.  This is not the case when the
system is out of equilibrium.  The aging regime is that in which
one-time quantities (such as energy or density) are stationary or
change very little, but two-time quantities depend explicitly on both
times $t_w, t$ \cite{revaging}.

An important quantity which allows to distinguish among different
kinds of aging behaviour \cite{revaging} is the
fluctuation-dissipation (FD) ratio,
\begin{equation}
X = T {R(t,t_w) \over \partial C(t,t_w) / \partial t_w } \ .
\label{FDR}
\end{equation}
In equilibrium, the fluctuation-dissipation theorem (FDT) states that
$X=1$, while out of equilibrium the FDT is violated and $X$ depends on
$t,t_w$ in a non trivial way.  It has been conjectured \cite{ck} that
during aging $X$ depends on time only through the correlation
function, conjecture that has proved valid in all systems studied to
date.  If one considers the susceptibility, or integrated response,
\begin{equation}
 \chi(t,t_w) = \int_{t_w}^t R(t,s) \, ds ,
\end{equation}
which is much easier to compute numerically, then assuming
$X=X[C(t,t_w)]$ one finds from (\ref{FDR})
\begin{equation}
 \chi(t,t_w) = \beta \int_{C(t,t_w)}^{C(t,t)} X(C) \, dC.
\end{equation}
Therefore one can plot $\chi$ vs.\ $C$ at fixed $t_w$, and extract $X$
from the slope of the curve. In equilibrium $X=1$ and  the curve is
a straight line $\chi(t-t_w)= \beta [C(0)-C(t-t_w)]$, with slope
$-1/k_BT$. For a system out of equilibrium one finds in general two regimes:
 a short time quasi-equilibrium regime ($t-t_w \ll t_w$), corresponding to
equilibration of fast degrees of freedom, where the curve follows the 
equilibrium straight line and FDT holds; and the aging regime at larger times  
($t-t_w \ll t_w$) where $\chi(t,t_w)$ departs from the equilibrium line.
In particular, for structural glass models studied so far via numerical 
simulations  \cite{kob-barrat,giorgio-aging} 
it has been found that after leaving the equilibrium line, the curves
follows another straight line, with smaller (in absolute value) slope.

To study the FD relations in CTLS we follow (as has been done before
in other spin systems \cite{alain}) the evolution of the staggered
magnetization ($\eta_i=\pm 1$ are quenched random variables),
\begin{equation}
m_s(t) = {1\over N} \sum_i \eta_i s_i(t) ,
\end{equation}
when adding a term $-h N m_s$ to the Hamiltonian. If the field $h$ is
applied at time $t_w$, the $\eta$-averaged susceptibility
${\chi(t,t_w)} = \overline{ m_s(t)}/h $ is related to the
spin-spin correlation function:
\begin{equation}
C_s(t,t_w) = {1\over N} \sum_i \langle s_i(t_w) s_i(t) \rangle.
\end{equation}
We have used fields $h=0.1\, k_B T$ and $h=0.15 \,k_B T$, and checked
that they are within the linear regime. The results presented here
correspond to the highest field.

\begin{figure}
\begin{center}
\leavevmode
\epsfxsize=3.5in
\epsffile{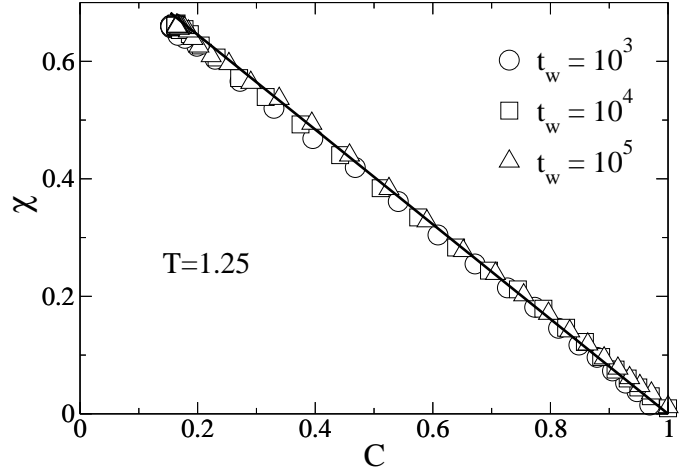}
\vglue 0.2 truecm \caption{ Parametric plot of $\chi(t,t_w)$ vs $C(t,t_w)$, at various
values of the waiting time $t_w$ for $T=1.25$, above the spinodal
temperature. $L=500$.  The full line represents the equilibrium relation
$\chi=\beta(1-C)$.}
\label{FDT1}
\end{center}
\end{figure}

We look for aging behavior in the regions $\Tsp < T < \Tm$ and $T <
\Tsp$. As we have seen, in both these regions the system has trouble
equilibrating, though for different reasons. We first quench the
system to $T=1.25$, which is above the spinodal, and plot
${\chi}$ vs.\ $C_s$ for different waiting times
(Fig.~\ref{FDT1}). Except perhaps for the shortest $t_w$, the curve
obeys the FDT: we obtain a straight line with slope $-1/k_B T$,
independent of $t_w$. No aging is observed, and the behavior is
compatible with a system in equilibrium, except that we know from
thermodynamics that the equilibrium phase at this temperature is the
crystal: we are witnessing the {\em metastable liquid,} which at this
temperature lasts much longer than its relaxation time, and can thus
be observed.

\begin{figure}
\begin{center}
\leavevmode
\epsfxsize=3.5in
\epsffile{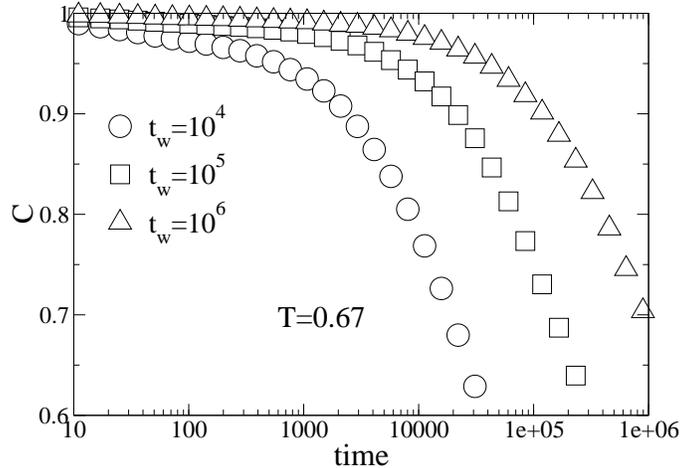}
\vglue 0.2 truecm \caption{Correlation function as a function of
$t$, at different values of the waiting time $t_w$. $T=0.67$,
$L=500$.}
\label{FDT2}
\end{center}
\end{figure}

\begin{figure}
\begin{center}
\leavevmode
\epsfxsize=3.5in
\epsffile{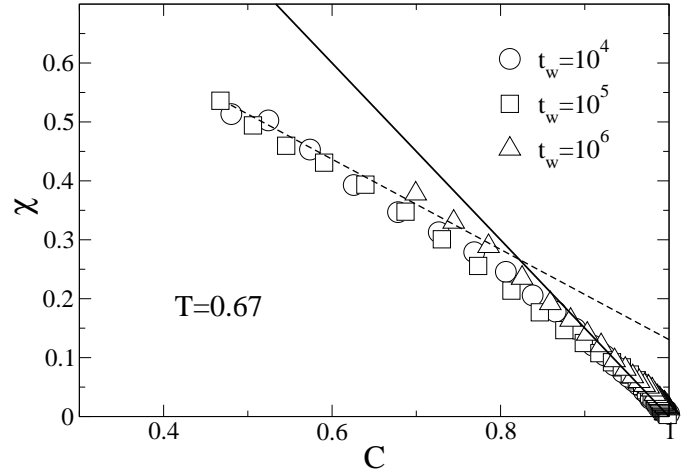}
\vglue 0.2 truecm \caption{Parametric plot of $\chi(t,t_w)$ vs $C(t,t_w)$, at various
values of the waiting time $t_w$ for $T=0.67$, well below the spinodal
temperature. $L=500$.  The full line represents the equilibrium relation
$\chi=\beta(1-C)$}
\label{FDT3}
\end{center}
\end{figure}

On the other hand, a quench to $T=0.67\ll \Tsp$ shows aging very clearly, as
the correlation function plotted for different waiting times shows
(Fig.~\ref{FDT2}). 
The FD ratio (Fig.~\ref{FDT3}) also shows very
clearly that the system is out of equilibrium. In fact, we know that
the liquid is unstable at this temperature, and that the crystal phase
is growing.  In a system undergoing coarsening, like the Ising model
quenched below $T_c$, the susceptibility rapidly saturates at the
value corresponding to the low-temperature phase, because the
contribution of the interfaces is negligible. The correlation
function, however, continues to drop because of the movement of the
domain walls. Thus the FD curve is flat for small values of the
correlation function \cite{cuku-peliti,alain,early}.  We would expect
this same behavior whenever the dynamics is given by growing domains
with bulk equilibrium properties, separated by sharp interfaces.
However, the FD curve does not become flat in the CTLS at $T=0.67$,
even for our longer waiting times. Instead, the
system shows FDT violations similar to those commonly exhibited by
structural glass models \cite{giorgio-aging,kob-barrat} and certain
mean field spin-glasses (solved by a one step replica broken solution,
1-step RSB) \cite{ck}. There  are two regimes: a quasi equilibrium one with
slope $-\beta$ and an off-equilibrium one with slope $-\beta_{\mathrm
eff}=-X \beta$. Note that the FD curves hardly depend on $t_w$, at
least for the times the simulations can reach.

Taken at face value, this result would lead to interpret this as an
out-of-equilibrium glassy phase (i.e.\ a liquid with weakly-broken
ergodicity). To understand it correctly however we need to recall the
discussion of the previous section.  We have seen that at low
temperatures ($T<\Tsp$) the crystal growth dynamics proceeds through
two qualitatively different time regimes. In the first bubbling regime
disordered regions separates small crystal domains and the interfacial
contribution to the susceptibility is still very large.  The
coarsening regime, where the dynamics is dominated by the nucleated
regions and the interfacial contribution can be neglected, will only
be reached later, when the crystal regions have grown enough to
overwhelm the interfaces.  Due to the extremely slow activated
dynamics the bubbling regime can last for long, and indeed at very low
temperatures it is the only one accessible in our simulation times. If
this is the case, we do not expect to see in the response-correlation
relation the FDT violations typical of coarsening systems, since the
coarsening regime has not yet been attained.  
This is exactly what
happens for $T=0.67$: we can check that the crystal mass for the
longest time explored in Fig.~\ref{FDT3} is $m \sim 0.3$
(fig. ~\ref{mass-xi}-Right), well below the value $m\sim 0.5$ that
characterizes the crossover to the coarsening regime. The 1-RSB-like
FDT violation observed at $T=0.67$ thus occurs in the bubbling regime,
while if we could wait longer times we would observe a crossover to 
typical a coarsening behaviour.

\vskip 0.5 cm

\begin{figure}
\begin{center}
\leavevmode
\epsfxsize=3.5 in
\epsffile{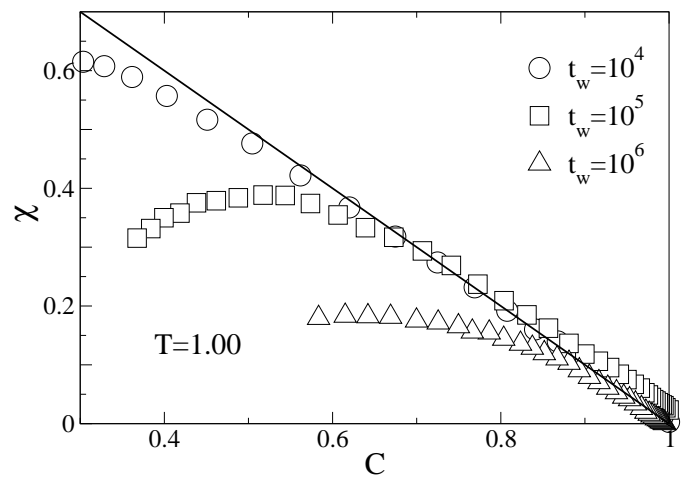}
\vglue 0.2 truecm \caption{
Parametric plot of $\chi(t,t_w)$ vs 
$C(t,t_w)$, at various values of the waiting time $t_w$ for
$T=1.00$, below the spinodal temperature. $L=500$.  
The full line represents the equilibrium relation $\chi=\beta(1-C)$.}
\label{FDT4}
\end{center}
\end{figure}

Since the duration of the bubbling regime is shorter for higher
temperatures, we expect to explicitly observe an FDT crossover to the
coarsening regime at higher temperatures. From the former Sections we know
that at $T=1.00$ the crossover from the bubbling to the coarsening regime
takes place at about $10^4-10^5$ MCS. In Fig.~\ref{FDT4} we plot the FD curves 
for $T=1.00$, and the data confirm our expectations.
For short waiting times, the curves initially follow what
seems to be the path of a 1-RSB-like violation. This is the bubbling 
phase, pretending to be a glass. The curves depend strongly on waiting time, 
and for higher $t_w$ evolve toward the normal domain-growth curve 
(two straight lines, with slope $-\beta$ at short times and 0 at long times), 
which is nearly reached for $t_w=10^6\,$MCS.

\vskip 0.5 cm 

\section{Conclusions}

We have presented a lattice model, the deterministic CTLS, which
exhibits a phenomenology typical of glass forming systems: it has a
melting transition and a low temperature crystalline ground state, but
it exists as a supercooled liquid for a wide range of temperatures; in
the metastable supercooled phase the relaxation time increases in a
way resemblant of typical fragile glasses and upon fast cooling
crystallization can be easily avoided to bring the system in
disordered glassy-like states.

The feature that makes this model particularly interesting is that it
displays a metastability limit which can be explicitly observed within
the time scales available for numerical simulations. What we have
shown is that the typical phenomenology of fragile glass forming
systems is entirely compatible with the existence of a metastability
limit of the supercooled phase, even if it is plausible that,
contrarily to the CTLS, for many models this limit cannot be observed.
A crucial point in this regard concerns the length of experimental
times under which measurements are performed. By definition the
supercooled liquid looses stability at $\Tsp$, when crystal nucleation
becomes faster than liquid equilibration.  If the equilibration time
at $\Tsp$ is much larger than the experimental time, however, the
metastability limit cannot be observed. Since the glass temperature
$T_g$ is the lowest temperature at which the system is equilibrated
within the experimental time, this situation takes place when
$\Tsp<T_g$.

If the system has a metastability limit, below the pseudospinodal
$\Tsp$ the dynamical behaviour basically consists of very slow crystal
growth.  This suggests some criteria to detect whether a system has
such a metastability limit or not, even when $T_{\rm sp}$ is
experimentally inaccessible.  For example, one would {\it a priori}
think that there is a qualitative difference between a disordered
glassy configuration, obtained by quenching a liquid, and a
crystalline configuration, however rich in defects this is, and
however slow crystal growth may be.  Also, since off-equilibrium
dynamics of simple domain growth has very peculiar characteristics one
would hope to detect crystal growth by analyzing the dynamical
behaviour of the system, and in particular the pattern of FDT
violations. What we have shown in the CTLS is that actually neither of
these criteria is sharp enough, and if our experimental time were not
long enough to {\it explicitly} observe the loss of stability of the
liquid at $T_{\rm sp}$, it would be impossible to make statements
about it, either by structural or dynamical means. 
On the other hand, on shorter time scales the system
exhibits an off-equilibrium behaviour which is typical of
structural fragile glasses. Indeed, we have seen that there is a long
time regime where FDT violations are 1-RSB like, as in most of fragile
glass models studied so far, while the proper coarsening regime is
attained only at much larger times. Also, there is structural
continuity between the low-temperature configurations reached via fast
coolings and those approached with slower cooling rates, which means
that the degree of order in the system increases gradually: what
appears as a disordered glassy like configuration is just the
beginning of a continuous process that eventually, on much longer time
scales, leads to the crystal.  The reason for this behaviour is that
below $T_{\rm sp}$ crystal nucleation is fast, but crystal growth
becomes very slow, with many crystal droplets trying to expand in a
liquid background.  In such a situation distinguishing between a truly
disordered glass and a mixture of tiny mismatched crystallites becomes
very hard, and FDT violation is nontrivial.

An important consequence of the metastability limit scenario in the
CTLS is that the Kauzmann paradox is avoided: below $\Tsp$ the liquid
does not exist in any reasonable sense, thus extrapolation of the
excess entropy is meaningless. This was the resolution of the paradox
proposed by Kauzmann himself in his paper.  While we cannot claim that
this is the case for real glasses, (and we have discussed qualitative
alternative scenarios in terms of crystal nucleation and growth), this
system shows that Kauzmann's way out of the Kauzmann paradox ought to
be considered seriously.  In the study of glassy systems the crystal
phase is often neglected, assuming that if crystallization is avoided
at $T_m$, then the crystal does not play any role in the low
temperature physics.  What this model shows is that the stability of
the supercooled liquid should not be taken for granted.

Finally, even if stability were to be proved not to be a concern in
real systems, this remains a cautionary note when formulating
non-disordered glass models, especially on the lattice, which is so
attractive.  One can also wonder if these considerations might be
applicable to disordered systems, or systems with complex (highly
degenerate) low-temperature phases, were the supercooled liquid could
be unstable with respect to a crystal in disguise.

\vskip 1 cm

We thank V.~Martin-Mayor and F.~Ricci-Tersenghi for many important
suggestions.  We also acknowledge interesting discussions with
P.~Debenedetti, H.~Horner, G.~Parisi, F.~Sciortino, D.~Sherrington and
P.~Verrocchio.  AC thanks in particular M.~A.~Moore.

\end{document}